\documentclass[reprint,nofootinbib,prl]{revtex4-1} 

\usepackage{amsmath,amssymb}
\usepackage{xcolor}
\usepackage{graphicx}
\usepackage{multirow}
\usepackage{dsfont}
\usepackage[braket,qm]{qcircuit}
\usepackage{booktabs,makecell}
\usepackage{cancel}
\usepackage{mathtools}   
\usepackage{siunitx}     
\usepackage{physics}
\usepackage[capitalize]{cleveref}

\newcommand{\mcdot}{\! \cdot \!}
\newcommand{\bn}{{\bar n}}

\newcommand{\ground}{{\Omega}}

\DeclarePairedDelimiter\ceil{\lceil}{\rceil}
\DeclarePairedDelimiter\floor{\lfloor}{\rfloor}

\providecommand{\oper}{\mathcal{Y}}
\providecommand{\ord}{\mathcal{O}}

\renewcommand{\vec}[1]{\ensuremath{\vb{#1}}}		
\providecommand{\gvec}[1]{\ensuremath{\vb*{#1}}}	

\begin{document}
\title{Simulating collider physics on quantum computers using effective field theories}

\author{Christian W. Bauer}
\email{cwbauer@lbl.gov}
\affiliation{Physics Division, Lawrence Berkeley National Laboratory, Berkeley, CA 94720, USA}
\author{Marat Freytsis}
\email{marat.freytsis@rutgers.edu}
\affiliation{NHETC, Department of Physics and Astronomy, Rutgers University, Piscataway, NJ 08854, USA}
\affiliation{Physics Division, Lawrence Berkeley National Laboratory, Berkeley, CA 94720, USA}
\author{Benjamin Nachman}
\email{bpnachman@lbl.gov}
\affiliation{Physics Division, Lawrence Berkeley National Laboratory, Berkeley, CA 94720, USA}

\begin{abstract}
Simulating the full dynamics of a quantum field theory over a wide range of energies requires exceptionally large quantum computing resources. Yet for many observables in particle physics, perturbative techniques are sufficient to accurately model all but a constrained range of energies within the validity of the theory. We demonstrate that effective field theories (EFTs) provide an efficient mechanism to separate the high energy dynamics that is easily calculated by traditional perturbation theory from the dynamics at low energy and show how quantum algorithms can be used to simulate the dynamics of the low energy EFT from first principles. As an explicit example we calculate the expectation values of vacuum-to-vacuum and vacuum-to-one-particle transitions in the presence of a time-ordered product of two Wilson lines in scalar field theory, an object closely related to those arising in EFTs of the Standard Model of particle physics. Calculations are performed using simulations of a quantum computer as well as measurements using the IBMQ Manhattan machine.
\end{abstract}
\maketitle


It is well known that quantum computers can in principle simulate the time evolution of quantum field theories~\cite{Jordan:2011ne}. 
The main technique involves disretizing the spatial degrees of freedom by introducing a lattice~\cite{Kogut:1974ag, Kogut:1979wt, Bronzan:1984xb}, and digitizing the field values at a given lattice point~\cite{Jordan:2011ci, Macridin:2018gdw, Macridin:2018oli, Klco:2018zqz, Hackett:2018cel, Yeter-Aydeniz:2018mix, Kreshchuk:2020dla, Kreshchuk:2020kcz, Haase:2020kaj}.
This turns the uncountably infinite dimensional Hilbert space of standard quantum field theories into a finite dimensional Hilbert space of dimension
\begin{align}
  n_H =n_\phi^{\pqty{N^d}}\,,
\end{align}
where $n_\phi$ denotes the dimensionality of the Hilbert space for a given lattice point, $N$ is the total number of lattice points in each spatial direction, and $d$ represents the number of spatial dimensions. 
The physical volume of the lattice is determined by the distance between adjacent lattice points $\var{x}$ in each direction and is given by
\begin{align}
  V = (N\var{x})^d \equiv L^d \,.
\end{align}
%
The total number of qubits required for such a simulation is given by
\begin{align}
  n_Q = \ord(\log_2 n_H) = \ord(N^d \log_2 n_\phi) \,.
\end{align}

The discretization and finite volume of space introduce upper and lower cutoffs to the energies $E$ over which the resulting lattice field theory is a good approximation for the continuum. 
In particular, one finds
\begin{align}
  \frac{1}{N\var{x}} \lesssim E \lesssim \frac{1}{\var{x}} \,,
\end{align}
which implies that the range of energies that can be described is directly proportional to the number of lattice sites per dimension.
In principle, to have access to the full dynamics of the Large Hadron Collinder (LHC), one would need to describe the energy range between $\ord(\SI{10}{\MeV})$ (the smallest resolvable transverse momentum between hadrons) and \SI{7}{\TeV} (the beam energy of the LHC).
To fully capture this energy range would require a lattice with $\ord(\num{e6})$ lattice points in each dimension, and more than \num{e18} qubits to reproduce the resulting physical system.
Even if one does not require the full energy range up to the LHC center-of-mass energy, the number of qubits required for a full simulation will clearly remain beyond the realm of feasibility for a long time to come.

For many observables of interest at the LHC, physics at short distances is reliably computed in fixed order perturbation theory, and high precision can be reached with existing techniques.
Physics at lower energies introduces new significant challenges.
Asymptotic freedom~\cite{Gross:1973id,Politzer:1973fx} implies that the strong coupling constant becomes large at low energies, increasing the production of additional particles.
An energetic particle can thus easily radiate additional soft and collinear particles, resulting in a collimated jet of particles. 
This means that fixed order perturbative calculations are of little use to describe lower energy effects such as the precise makeup of jets, and other techniques such as resummation~\cite{Collins:1984kg,Catani:1992ua,Bonciani:2003nt} and parton showers~\cite{Marchesini:1983bm,Bengtsson:1986et} have to be used to make predictions.
At even lower energies, the fully nonperturbative dynamics of hadronization dominate.
While classical lattice methods provide spectral information about the strongly-coupled limit of the theory, the dynamics of hadronization is currently only understood through phenomenological models and form factors extracted from data.
While these techniques have been quite successful and it has even been shown that quantum algorithms can be used to include quantum interference effects in some models of parton showers~\cite{Bauer:2019qxa}, it would be a significant breakthrough were it possible to simulate these low energy dynamics from first principles.

In this paper we show that this is indeed possible by using effective field theories (EFTs) which have been designed to reproduce the desired long distance physics.
(For work on simulating EFTs not related to collider physics, see~\cite{Dumitrescu:2018njn,Lu:2018pjk,Cervia:2019res,Roggero:2019myu,Cervia:2020fkk,Stetina:2020abi}).
The dynamics of these EFTs can then be simulated directly and from first principles on a quantum computer. 
Since the energy range that needs to be simulated is much smaller than that of the full theory, the resource requirements are orders of magnitudes smaller than those required to simulate the full theory.
Additionally, the EFT simplifies the description of the initial and final state and significantly reduces the resource requirements of their implementation.
For example, consider an observable measured on two jets, each with an energy of $\ord(\SI{1}{\TeV})$ and an invariant mass of $\ord(\SI{50}{\GeV})$. 
Even restricting ourselves to observables insensitive to hadronization, a generic observable of the momenta of final state particles in the full theory requires a simulation of the range $\SI{1}{\GeV} \lesssim E \lesssim \SI{2}{\TeV}$.
As we discuss below, the EFT only needs to capture $\SI{1}{\GeV} \lesssim E \lesssim \SI{50}{\GeV}$.
The number of qubits required to simulate the EFT is smaller by a factor of $(2000/50)^3 \approx \num{e5}$, and this factor increases rapidly as the range of the full theory is increased.

The EFT relevant for hadronic jet physics is Soft-Collinear Effective Theory (SCET)~\cite{Bauer:2000ew,Bauer:2000yr,Bauer:2001ct,Bauer:2001yt}. 
Using this EFT, a typical cross section describing a physical observable at the LHC can be factorized into separate pieces~\cite{Bauer:2002nz,Bauer:2002ie,Bauer:2003di,Manohar:2003vb}
\begin{align}
  \sigma = H \otimes J_1 \otimes \dotsb \otimes J_n \otimes S \,.
\end{align}
Here $H$ denotes a coefficient describing the short distance physics which can be computed reliably in standard perturbation theory, while $J_i$ and $S$ denote squares of matrix elements of operators in SCET.
The jet function $J_i$ denotes physics arising from collinear degrees of freedom that all have small momentum relative to each other (while moving collectively with a large energy), with the different jet functions completely decoupled from one another.
The soft function $S$ denotes physics arising from soft degrees of freedom, which all have small absolute momentum (in the frame of the collider).
These matrix elements describe the transition of a simple initial state produced in the short distance interaction $H$ into a final state, with the dynamics described by the EFT Hamiltonian.
The final states that arise can contain a large number of particles.
State of the art techniques use operator renormalization to compute the overall scale dependence of the jet and soft functions, and then compute the relevant matrix elements perturbatively, such that the effects arising from the high multiplicity final states or hadronization can not be properly included.
Classical lattice techniques are also not suitable for the computation of matrix elements of SCET operators, since the long-distance dynamics is governed by massless modes which are inherently Minkowskian in nature.
Having a simulation of the dynamics of SCET (or a different EFT for other problems) will allow for a full non-perturbative calculation of any matrix elements.

The EFT reproduces the full theory result up to power corrections, whose size depends on the kinematics of the process studied. 
For many observables of interest, these power corrections are considerably smaller than the perturbative corrections present in either the full theory or EFT matrix elements.

Since individual jet functions do not interact with one another, their dynamics can be simulated in the reference frame where all particles have small absolute momentum~\cite{Becher:2006qw,Goerke:2017ioi}.
This requires simulating the dynamics of the original field theory (with any degrees of freedom that only contribute to short distance effects removed), but with a much smaller energy range required to be simulated. 
To achieve this on a Quantum Computer, one needs a setup very similar to that for the full theory, but reliable calculations can be achieved with much coarser lattices than those one would need for a simulation of the full theory.

In the soft function, on the other hand, the energetic particles no longer contribute to the dynamics of the theory.
Instead, their effect is captured by a so-called Wilson line, which describes the interaction of a charge moving along a fixed world-line with the bath of soft degrees of freedom.
The physics underlying this observation is that soft particles cannot change the direction of energetic particles in a meaningful way. 
This implies that energetic particles can be included in the EFT as a static object (Wilson line) described by the relevant quantum numbers (charge, color,  etc.) moving along a fixed world line along a light-like direction.
Matrix elements in the soft theory therefore need to compute the dynamics of a Hamiltonian describing the soft bath of particles in the presence of such Wilson lines.
For gauge theories, such as those describing the three fundamental forces contained in the Standard Model, Wilson lines are given by path ordered exponentials of the relevant gauge fields~\cite{Bauer:2001yt}
\begin{align}
  Y_n = \mathrm{P} \exp\bqty{ i g \int_0^\infty \!\!\dd{s} \, n\mcdot A(x^\mu = n^\mu s) } \,.
\end{align}
Here $A^\mu$ is the soft gauge field and $n^\mu$ describes the direction of the light-like direction $n^\mu = (1, \vec{n})$ with $\vec{n}^2 = 1$ such that $n_\mu n^\mu = 0$. 
Thus, the gauge field is evaluated on a path going from the origin to spatial infinity along the world-line characterized by the direction $n^\mu$. 
The dynamics of the soft gauge field is described by its full theory Hamiltonian.

The soft function for a process containing two energetic particles of zero total charge moving back-to-back requires two Wilson lines, $Y_n$ for the particle moving in the $n$ direction and $Y_\bn^\dagger$ for the antiparticle moving in the $\bn^\mu = (1, -\vec{n})$ direction. 
The matrix elements required in the soft sector are given by
\begin{align}
\label{eq:WilsonMatrix}
  \mel{X}{ \mathrm{T}[Y_n \, Y_{\bn}^\dagger] }{\ground} \,,
\end{align}
where $\mathrm{T}$ denotes the time-ordering operator, $\ket{\ground}$ denotes the ground state of a Hilbert space containing only the gauge degree of freedom, and $\ket{X}$ is the final state.
If reproducing multi-particle weakly coupled final states, $\ket{X}$ will contain a fixed number of gauge momentum modes with given momenta $k^\mu_i$.
To compute the soft function for a given observable one needs to sum over all possible final states that can contribute to this observable.
Traditional perturbative calculations~\cite{Becher:2005pd,Hoang:2008fs,Jouttenus:2011wh,Kelley:2011ng,Monni:2011gb,Boughezal:2015eha,Moult:2018jzp} can only compute these matrix elements for final states $\ket{X}$ containing a small number of particles and at low order in perturbation theory, since the complexity of the calculation increases factorially with the power of the coupling constant $g$. 
They therefore do not compute the full matrix element for a given observable, but only a perturbative approximation.
For large coupling constants, which is the relevant case for Quantum Chromodynamics (QCD) at low energies, such perturbative calculations can give rise to large uncertainties.

Simulating the full dynamics of the field theory would provide the full non-perturbative result for the soft matrix element. 
To evaluate the matrix element on a quantum computer one first needs to define circuits that perform time evolution of the system, as well as circuits that can create and measure the ground and exited states of the theory. 
In addition, one needs to create circuits that can correctly interleave the implementation of nonlocal Wilson line operators with the evolution of the system to reproduce their time-ordered product.

In the following we discuss how to compute matrix elements of Wilson line operators $Y_n$ analogous to \cref{eq:WilsonMatrix}, but for a massless scalar, rather than gauge, theory.
The EFT is insensitive to the precise origin of the Wilson lines, but a particualy straightforward realization would result from a pair of highly-energetic fermions coupled to massless scalars through a Yukawa coupling.
When constructing the explicit circuit we also limit ourselves to (1+1) dimensions, mainly to restrict the quantum resources required such that it can be implemented on currently existing hardware.
This allows us to omit some technical complications that arise when dealing with gauge theories (gauge transformations, the existence of unphysical polarizations, \emph{etc}.), while capturing all the resulting simplification of working within an EFT.

To be precise, we consider a massless field theory in (1+1) dimensions with Hamiltonian and Wilson lines defined by
\begin{align}
  H &= \int \dd{x} \frac{1}{2} \pqty{ \dot{\phi}^2 - \phi \, \partial^2 \phi } \,\qc
  \nonumber\\
  Y_n & = \mathrm{P} \exp \bqty{ ig \int_0^\infty \!\!\dd{s} \, \phi(x^\mu = n^\mu s) } \,.
\end{align}
In the Supplemental Material, we discuss how working in (1+1) dimensions gives rise to several effects not present in higher dimensions, and how these generalize to higher dimensions.

We discretize the position $x$ into an odd number of lattice points, labeling the positions by $x_{0}, \dotsc,  x_{N-1}$. 
To eliminate the zero-momentum mode of the theory, we impose twisted boundary conditions~\cite{Lin:2001zz,Sachrajda:2004mi,Bedaque:2004kc,Briceno:2013hya}.
The result is a theory defined at discrete positions $x$ and momenta $p$ given by $x_i = x_\mathrm{min} + i\var{x}$ and $p_i = p_\mathrm{min} + i\var{p}$ with $x_\mathrm{min} =  -(N - 1)\var{x}/2$, $p_\mathrm{min} = -\pi / \var{x}$ and $\var{p} = 2\pi / N \var{x}$, 
Writing $\phi_i \equiv \phi(x_i)$, the twisted boundary conditions correspond to the condition $\phi_{i+N} = -\phi_i$. 
The Hamiltonian becomes~\cite{Jordan:2011ci}
\begin{align}
  H = \frac{\var{x}}{2} \sum_{i=0}^{N-1} \bqty{ \dot{\phi}_i^2
                                      - \phi_i \, [\nabla^2 \phi]_j } \,,
\end{align}
where the lattice operator $\nabla^2$ is defined through its action on a field as $[\nabla^2 \phi]_i = (2 \phi_i - \phi_{i-1} - \phi_{i+1}) / \var{x}^2$. Due to the twisted boundary conditions $[\nabla^2 \phi]_0 = (2 \phi_0 + \phi_{N-1} - \phi_{1}) / \var{x}^2$ and $[\nabla^2 \phi]_{N-1} = (2 \phi_{N-1} - \phi_{N-2} + \phi_{0}) / \var{x}^2$. 
The Wilson line operators can be written as
\begin{align}
  Y_n &= \mathrm{P} \exp\bqty{ ig \var{x} \sum_{i=n_0}^{2n_0}\phi_{x_i}(t = x_i - n_0) } \,\qc
  \nonumber\\
  Y_{\bn}^\dagger
    &= \mathrm{P} \exp\bqty{- i g \var{x} \sum_{i=0}^{n_0}\phi_{x_i}(t = n_0 - x_i)} \,,
\end{align}
where $n_0 = (N-1)/2$ denotes the point at the center of the lattice.

We represent the field theory through the field values at each lattice position, and in order to describe the theory on a digital quantum computer one needs to digitize the continuous field value at each lattice point~\cite{Klco:2018zqz}.
Choosing $n_Q$ qubits per lattice site allows for $n_\phi \equiv 2^{n_Q}$ different field values.
For each lattice point, the possible field values are chosen to be by $\phi_i^{(k)} = -\phi_\mathrm{max} + k\, \delta \phi$, with $\delta \phi = 2 \phi_\mathrm{max} / (n_\phi - 1)$. 
The value of $\phi_\mathrm{max}$ has to be chosen to optimize the digitized description, which for free fields is accomplished by
\begin{align}
\label{phiMax}
  \phi_\mathrm{max} = \frac{1}{\sqrt{\delta x \, \bar{\omega}}}
                        \sqrt{\frac{\pi}{2} \frac{(n_\phi-1)^2}{n_\phi}}
                        \,,
\end{align}
where 
\begin{align}
\label{omega}
  \bar{\omega} = \frac{1}{N} \sum_i \omega_i \,\qc
  \omega_i = \frac{2}{\var{x}} \abs{ \sin \frac{p_i \var{x}}{2} } \,.
\end{align}
For $\bar{\omega} = 1$, as is the case for a single lattice site with $\omega = 1$, corresponding to a single harmonic oscillator, \cref{phiMax} reproduces the empirical numerical values obtained in~\cite{Klco:2018zqz}. 

To implement the Wilson line operator we first rewrite the time-ordered product of the two Wilson lines as
\begin{align}
\label{WilsonLineOp}
  \mathrm{T}[Y_n \, Y_\bn^\dagger]
    &= e^{ -iH\, n_0\var{x}} \exp\bqty{ ig \var{x} \pqty{ \phi_{x_{2n_0}} - \phi_{x_0} }} 
    \\
    & \qquad  \times    e^{ iH \var{x}} \exp\bqty{ i g \var{x} \pqty{ \phi_{x_{2n_0-1}} - \phi_{x_1} }}\nonumber\\
    &\qquad \times \dotsm \times
        e^{i H \var{x}} \exp\bqty{ i g \var{x} \pqty{\phi_{x_{n_0}} - \phi_{x_{n_0}} }} \,,\nonumber
\end{align}
where we have used the time translation operator to make the time dependence on the field operators explicit. 
Thus, the Wilson line operator consists of a sequence of time-evolution operators for a time interval corresponding to the lattice spacing and exponentials of the field operator. 
The last time evolution evolves the state back from the largest time to which the Wilson lines can be sensibly evolved, namely $t_\mathrm{max} = n_0 \var{x}$, to $t = 0$ at which all states are defined.

Ultimately, to make contact with the continuum field theory any such simulation will have to be performed on a series of increasing lattices, and the result extrapolated to the $N \to \infty$, $\var{x} \to 0$ limit.
Any parameters of the theory present in the continuum must be suitably matched for this procedure to yield meaningful results.
For local terms in the Hamiltonian, this procedure is discussed in detail in~\cite{Jordan:2011ci}.
Dealing with a massless theory simplifies this procedure since only local interactions (of which in the present case there are none) need to be matched.
However, the EFT will also require the matching of Wilson line operators, which is complicated by their non-local nature and sensitivity to total lattice size, as discussed in the Supplemental Materials.
In this letter, we work at fixed lattice size and we leave the detailed investigation of these issues to future work.

The implementation of the exponential of the field operator, as well as the time evolution operator, follows the discussion in~\cite{Klco:2018zqz} and uses the fact that the digitized field $\phi_i^{(k)}$ can be written in terms of sums of $\sigma_z$ operators.
This implies that the exponential of products of fields $\phi_i$ can be implemented through combinations of CNOT gates and $R_Z$ rotations~\cite{Macridin:2018gdw,Macridin:2018oli,Klco:2018zqz}. 
The exponential of the conjugate operator $\dot{\phi}^2$ can be implemented by taking a quantum Fourier transformation of the exponential of the operator $\phi^2$. 
These can then be combined via the Suzuki--Trotter formula~\cite{10.2307/2033649,Suzuki1976,1976PThPh..56.1454S}. 
For details, see the Supplemental Materials. 
The initial ground state of the scalar field theory is a multi-variate Gaussian distribution, which can be created using the approach of Kitaev and Webb (KW)~\cite{kitaev2009wavefunction}. 
To identify states of definite multiplicity and momentum in $\ket{X}$ one can follow the general ideas laid out in~\cite{Jordan:2011ne,Jordan:2011ci}.

Our quantum circuit has been implemented in \textsc{Qiskit}~\cite{Qiskit} and is available from the authors upon request.
In this first exploratory paper we compute the foundational quantities, namely 
\begin{align}
 \oper_X = \abs{ \!\mel{X}{\mathrm{T}[Y_n \, Y_{\bn}^\dagger]}{\ground} }^2
\end{align}
for $\ket{X} = \ket{\ground}$ and $\ket{X} = \ket{p_i}$, the one-particle momentum eigenstates of the theory.  
It should be noted that these quantities are not infrared (IR) safe, and will therefore depend on the IR scale in the problem, the lattice size $L$. 
However, as discussed in more detail in the Supplemental Materials, there is no non-trivial IR safe observable that can be defined in (1+1) dimensions, and these transition rates are therefore representative quantities of what can be computed in this theory.

The quantum circuit for this measurement can be represented as
\begin{align*}
\Qcircuit @C=0.5em @R=0.8em @!R{
  \lstick{\ket{l_0}}     & {/} \qw & \multigate{2}{U_\ground} &  \qw & \multigate{2}{U_Y}
    & \qw & \multigate{2}{U_X^\dagger} & \qw & \meter \\
  \lstick{\ket{\dotso}}  & {/} \qw  & \ghost{U_\ground}       &  \qw & \ghost{U_Y}
    & \qw & \ghost{U_X^\dagger}        & \qw & \meter \\
  \lstick{\ket{l_{N-1}}} & {/} \qw  & \ghost{U_\ground}       &  \qw & \ghost{U_Y}
    & \qw & \ghost{U_X^\dagger}        & \qw & \meter \\ 
}
\end{align*}
where $\ket{l_n}$ denotes the register of qubits for the $n^\text{th}$ lattice site. 
This creates the multivariate Gaussian vacuum state from the initial state with all qubits zero using $U_\ground$, acts on this vacuum with the time ordered product of the two Wilson lines using $U_Y$, and finally applies the inverse of the state preparation of state $\ket{X}$. 
The details of these various circuits can be found in the Supplementary Material.

\begin{figure}
  \centering
  \includegraphics[width=0.45\textwidth]{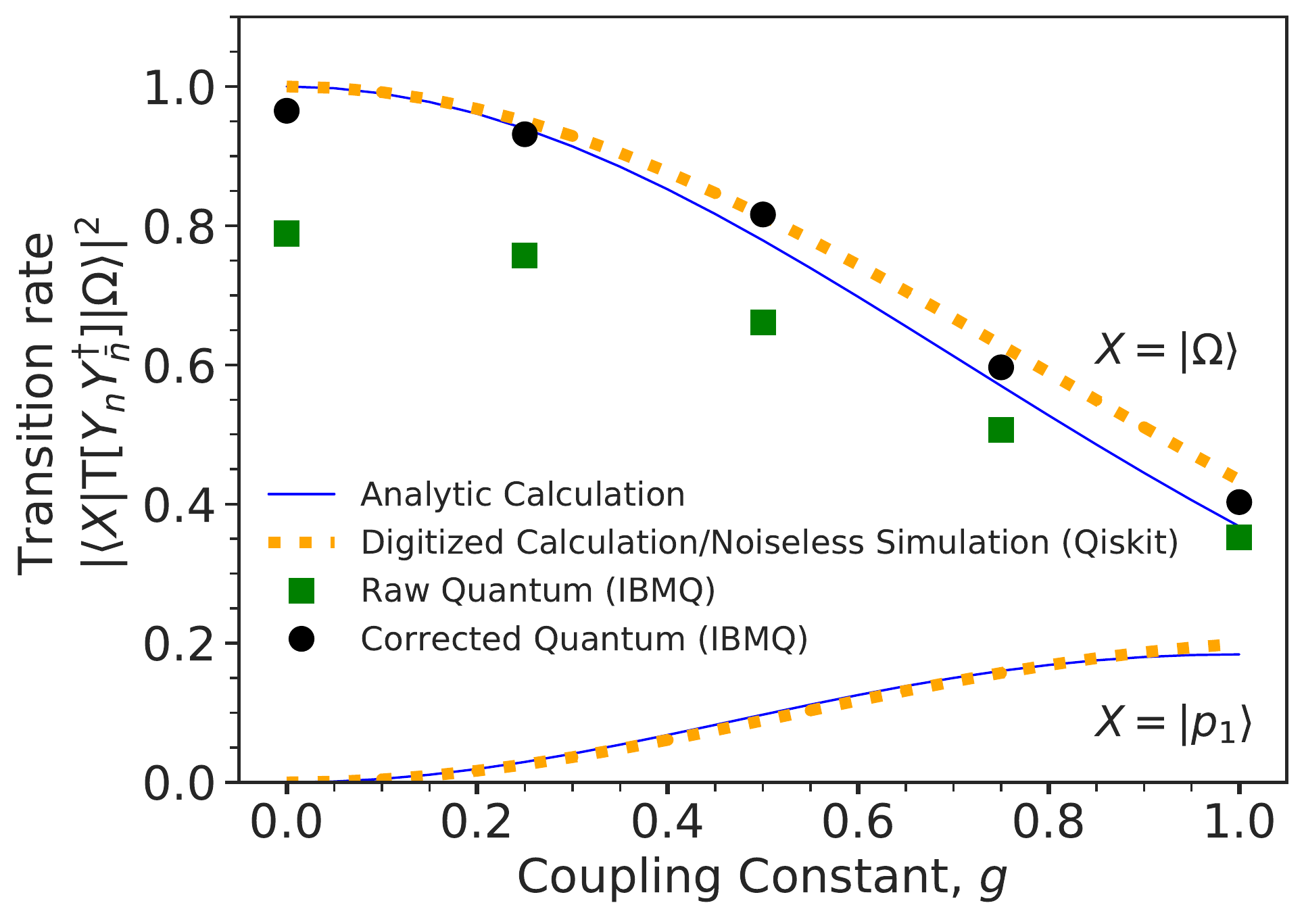}
  \caption{Result of transition rates from the valuum of the Wilson line for 3 lattice sites and $n_Q = 2$ qubits per site to the vacuum and the lowest-lying single excited state.  The solid lines shows the analytical result with no field digitization while the dashed lines represents the result from a quantum simulator of our circuit. The black data points show the result from the 65-qubit IBMQ Manhattan quantum computer, corrected both for readout errors and CNOT gate errors. We only show result from the Manhattan computer for $X = \ground$, since the circuit to measure the excited state was too deep to give reliable results.} 
\label{fig:vacExpectMelbourne} 
\end{figure}

For our numerical results, we work with an $N = 3$ site lattice with spacing $\var{x} = 1$.
With only 3 lattice sites, the Wilson line operator simplifies to 
\begin{align}
\label{WilsonLineOp3Lattice}
 \oper_X = \abs{ \!\mel{X}{\mathrm{T}[Y_n \, Y_{\bn}^\dagger]}{\ground} }^2 = \abs{ \!\mel{X}{e^{ig \var{x} \pqty{ \phi_{x_{2}} - \phi_{x_0} }}}{\ground} }^2
\,,\end{align}
since all time evolution operators act on the initial or final eigenstate of the Hamiltonian only and can therefore be neglected as contributing an overall phase.
In \cref{fig:vacExpectMelbourne} we show the dependence of the expectation values $\oper_X$ on the coupling $g$ for $n_Q = 2$ qubits per lattice site for different final states, and compare them against the analytical results, shown by black lines. 
Results are given for both a quantum simulation and from the 65-qubit IMBQ Manhattan quantum computer.
The operators for implementing all states are exact, as the resources for doing so on a small lattice are modest.
On a larger lattice approximate methods, such as KW ground state approximation and the excited state preparation techniques of~\cite{Jordan:2011ci} will be necessary; the effect of such approximations is presented in the Supplementary Material.

Errors in the quantum circuits, especially readout errors and CNOT gate errors are quite large on existing hardware. 
As discussed in the Supplemental Materials, the exponential of the field operator at a given position requires only $n_Q$ single qubit gates, such that the operator in \cref{WilsonLineOp3Lattice} requires no CNOT gates.
For $n_Q = 2$, he state preparation requires 6 CNOT gates for gates. 
Note that for more than 3 lattice sites the time evolution operator is required, which requires a much larger amount of gates, although the resulting circuits are known.
For example, even for 3 lattice sites the standard implementation of a single Trotter step of our Hamiltonian requires 60 CNOT gates.
We have applied both readout error mitigation as described in~\cite{ReadoutCorrection} as well as CNOT gate noise mitigation~\cite{He:2020udd}. For more details, including References~\cite{futureKW,DAgostini:1994fjx,1974AJ.....79..745L,Richardson:72,2010.07496,geller_efficient_2020,song_10-qubit_2017,gong_genuine_2019,hamilton2020scalable,wei_verifying_2020,pyquil,cirq,arute2020quantum,xacc,alex2019xacc,mccaskey_quantum_2019,ignis,1904.11935,1907.08518,Dumitrescu:2018,PhysRevX.8.031027,PhysRevLett.119.180509}, see the Supplemental Materials.
One can see that the digitized result with 2 qubits per lattice site differs from the analytic calculation by up to \SI{10}{\percent}. 
This would be reduced to at most \SI{1.5}{\percent} by adding just a single qubit per lattice site, since the resulting digitization errors fall exponentially with the number of qubits. 
The quantum computer reproduces the simulated result to about \SI{5}{\percent} accuracy.

In conclusion, EFT are well known to be able to describe the low energy dynamics of field theories and, through short distance, perturbatively computable matching coefficients, can be used to describe the dynamics of a full underlying quantum field theory.
We have argued that the dynamics of a low energy EFT can be simulated with significantly smaller quantum resources than the dynamics of the corresponding full theory. 
In SCET the interactions of highly energetic particles with soft particles of low energy are described through operators containing Wilson lines, and we have shown in detail how the dynamics of an analogous scalar soft theory can be described using quantum algorithms. 
Using Wilson lines of free scalar fields in (1+1) dimension, we have computed the simplest matrix elements in this soft theory, namely the transition matrix elements from the vacuum to itself and the lowest-lying excited states of two Wilson lines in opposite directions, using 3 lattice sites.
We have compared the computations on a quantum computer to analytical results that can be obtained for this simple theory. 
Using only 2 qubits per lattice we obtain results within \SI{10}{\percent} of the analytical result, and by using noise-mitigation techniques, uncertainties due to running on present-day hardware can be reduced to about \SI{5}{\percent}.

\section*{Acknowledgements}

We would like to thank Dorota Grabowska, Bert de Jong, Michael Kreshchuk, Pier Monni, John Preskill, Martin Savage and Miro Urbanek for useful discussions.  CWB and BN are supported by the U.S. Department of Energy (DOE), Office of Science under contract DE-AC02-05CH11231. In particular, support comes from Quantum Information Science Enabled Discovery (QuantISED) for High Energy Physics (KA2401032) and the Office of Advanced Scientific Computing Research (ASCR) through the Accelerated Research for Quantum Computing Program. MF is supported by the DOE under grant DE-SC0010008. This research used resources of the Oak Ridge Leadership Computing Facility, which is a DOE Office of Science User Facility supported under Contract DE-AC05-00OR22725.

\onecolumngrid

\clearpage
\newpage
\renewcommand{\theequation}{S\arabic{equation}}
\maketitle
\begin{center}
\textbf{\large Simulating effective field theories on quantum computers} \\ 
\vspace{0.05in}
{ \it \large Supplementary Material}\\ 
\vspace{0.05in}
{Christian W. Bauer, Marat Freytsis, Benjamin Nachman}
\end{center}

\setcounter{equation}{0}
\setcounter{figure}{0}
\setcounter{table}{0}
\setcounter{section}{0}
\setcounter{footnote}{0}
\setcounter{page}{1}


\section{Analytical calculations in the lattice field theory}

In this appendix we provide analytical calculations for the field theory under considerations that our quantum algorithms can be compared against.
The lattice field theory and its quantization agree with the analogous treatment in~\cite{Jordan:2011ci}, while the calculations involving the Wilson lines have not appeared anywhere else to our knowledge.

\subsection{Definition of the lattice}

We consider a lattice of $d$ dimensions, $N$ lattice sites for each dimension and lattice spacing $\var{x}$. Points on the latice are indexed by $d$ integers ranging from $0$ to $N-1$. 
We denote this set of integers by
\begin{align}
  \vec{r} = (r_1, \dotsc, r_d) \,\qc r_i \in \{ 0, \dotsc, N-1 \}
\end{align}
with the the positions of the lattice in each dimension given by
\begin{align}
\label{n0def}
  \vec{x}_{\vec{r}} = -\vec{x}_\textrm{max} + \vec{r}\var{x} \,\qc
  \vec{x}_\textrm{max} = ( n_0 \var{x}, \dotsc, n_0 \var{x} ) \,\qusing
    n_0 = \floor*{ \frac{N-1}{2} } \,.
\end{align}
The corresponding momentum values in each dimension are then given by (\vec{s} denotes the same set of integers as \vec{r})
\begin{align}
\label{pLatticeDef}
  \vec{p}_{\vec{s}} = -\vec{p}_\mathrm{max} + (\vec{s}-\gvec{\Delta}) \var{p} \,\qc \vec{p}_\mathrm{max} = ( n_0 \var{p}, \dotsc , n_0 \var{p} ) \,\qusing
    \var{p} = \frac{2\pi}{N \var{x}} \equiv \frac{2\pi}{L} \,.
\end{align}
Periodic boundary conditions correspond to $\gvec{\Delta} = (0, \dotsc, 0)$, while twisted boundary conditions (which are equal to periodic boundary conditions with an extra factor of $(-1)$ for each wrap around the lattice), correspond to $\gvec{\Delta} = (1/2, \dotsc, 1/2)$.

To denote the sum over all lattice positions or conjugate lattice momenta we define
\begin{align}
  \sum_{\vec{r}} \equiv \sum_{u=1}^d \sum_{r_u = 0}^{N-1} \,.
\end{align}
Note that the twisted boundary conditions ensure that the mode with vanishing momentum is not part of the spectrum. For each dimension every momentum has a partner with negative momentum
\begin{align}
  p_m = - p_{2n_0 - m + 1} \,\qfor m > 0 \,,
\end{align}
except for an odd number of lattice sites, in which case the momentum $p_0$ does not have corresponding partner. (Note that one can of course translate the lattice and conjugate lattice such that $\vec{x}_{\vec{r}} = \vec{r}\var{x}$ and $\vec{p}_{\vec{s}} = (\vec{s} + \gvec{\Delta}) \var{p}$.)

To go between fields defined on the position spaced lattice to those defined on the momentum spaced lattice one uses the discrete Fourier transform which is given by
\begin{align}
\label{FourierTrafo}
  \phi_{\vec{x}} = \frac{1}{(2\pi)^d}
                     \sum_{\vec{p}} e^{-i\vec{p}\cdot\vec{x}} \phi_{\vec{p}} \,\qc
  \phi_{\vec{p}} = \sum_{\vec{x}} e^{i\vec{p}\cdot\vec{x}} \phi_{\vec{x}} \,.
\end{align}
Note that we have introduced a short-hand form for the sum over lattice sites as
\begin{align}
  \sum_{\vec{x}} \equiv (\var{x})^d  \sum_{\vec{r}} \,\qc
  \sum_{\vec{p}} \equiv (\var{p})^d  \sum_{\vec{s}} \,,
\end{align}
where \vec{r} and \vec{s} denote the positions on the regular and conjugate lattice, respectively.

One has the usual completeness relation 
\begin{align}
  \frac{1}{(2\pi)^d} \sum_p e^{i \vec{p}\cdot(\vec{x}-\vec{x}')} = \delta_{\vec{x}\vec{x}'} \,\qc
  \sum_x e^{i \vec{x}\cdot(\vec{p}-\vec{p}')} = (2\pi)^d \delta_{\vec{p}\vec{p}'} \,,
\end{align}
were we have again used a short hand notation
\begin{align}
  \delta_{\vec{x}\vec{x}'} \equiv \frac{1}{(\delta x)^d} \delta_{\vec{r}\vec{r}'} \,\qc \delta_{\vec{p}\vec{p}'} \equiv \frac{1}{(\delta p)^d} \delta_{\vec{s}\vec{s}'} \,,
\end{align}
and we have chosen the factor of $2\pi$ to reflect the standard choice made in continuum field theory.

Before we conclude this section, we briefly discuss the relationship between the boundary conditions and the value of $\gvec{\Delta}$. 
From the definition of the Fourier transformation in \cref{FourierTrafo}, one can immediately see that transforming a field by the lattice size in some dimension yields
\begin{align}
  \phi_{\vec{x} + L \vec{n}_i} = e^{i (2 \pi\gvec{\Delta}_i)} \phi_{\vec{x}} \,.
\end{align}
Thus, non-integer values of $\gvec{\Delta}_i$ give an extra phase in the boundary condition, with half-integer values adding a relative minus sign, giving twisted boundary conditions.

\subsection{The free field theory}

The free scalar field theory on the lattice is given by the Hamiltonian
\begin{align}
  H = \frac{1}{2} \sum_{\vec{x}} \bqty{ \dot{\phi}_{\vec{x}}^2
        - \phi_{\vec{x}} \, [\grad^2 \, \phi]_{\vec{x}} } \,,
\end{align}
where the sums run over the $n$ possible integers (lattice sites) for each dimension. 
The d'Alembertian operator is defined through its action on a field, analogously to the second derivative operator in the main text.
Performing a Fourier transform, one can write this expression as
\begin{align}
  H = \frac{1}{2} \frac{1}{(2\pi)^d} \sum_{\vec{p}} \bqty{ \dot{\phi}_{\vec{p}} \dot{\phi}_{-\vec{p}}
        + \omega_{\vec{p}}^2 \, \phi_{\vec{p}} \phi_{-\vec{p}} } \,,
\end{align}
where frequencies are given by
\begin{align}
  \omega_{\vec{p}} = \frac{2}{\var{x}}
                       \sqrt{ \sum_{i=1}^d \sin^2 \pqty{\frac{\vec{p}_i \var{x}}{2}} } \,.
\end{align}
Introducing the creation and annihilation operators as
\begin{align}
\label{ladderopdef}
  \dot{\phi}_{\vec{p}} = -i \sqrt{\frac{\omega_{\vec{p}}}{2}}
                              \pqty{ a_{\vec{p}} - a^\dagger_{-\vec{p}} } \,\qc 
  \phi_{\vec{p}} = \frac{1}{\sqrt{2 \omega_{\vec{p}}}}
                     \pqty{a_{\vec{p}} + a^\dagger_{-\vec{p}} } \,,
\end{align}
which satisfy the commutation relations
\begin{align}
  \comm{ a_{\vec{p}} }{ a^\dagger_{\vec{q}} } = \pqty{\frac{2\pi}{\var{p}}}^d \delta_{\vec{p}\vec{q}} \,,
\end{align}
one finds for the Hamiltonian
\begin{align}
  H = \frac{1}{(2\pi)^d} \sum_{\vec{p}} \omega_{\vec{p}}
        \bqty{ a_{\vec{p}} a^\dagger_{\vec{p}} + \frac{1}{2}\pqty{\frac{2\pi}{\var{p}}}^d } \,.
\end{align}

Thus, in momentum space the free field theory on a lattice is simply a collection of uncoupled harmonic oscillators, and one can therefore compute the spectrum of the theory rather easily.
A general state is therefore labeled by its $d\times n$ occupation numbers $\vec{l}^{(0)} \dotsb \vec{l}^{(n-1)}$, where the occupation number at each lattice site is determined by $d$ integers $\geq 0$. Thus, a general state is given by
\begin{align}
  \ket{\Psi_{l_{\vec{k}}}} = \bigotimes_{l_{\vec{k}}} \ket{l_{\vec{k}}} \,,
\end{align}
with the ground state given by 
\begin{align}
  \ket{\ground} = \bigotimes_{l_\mathbf{k}} \ket{0} \,.
\end{align}

The energy of the ground state $\ket{\ground}$ is given by 
\begin{align}
  H \ket{\ground} = E_0 \ket{\ground} \,\qc
  E_0 = \frac{1}{2} \sum_{\vec{s}} \omega_{\vec{s}} \,,
\end{align}
while the energy of excited states are easily determined in terms of their occupation numbers as
\begin{align}
  H \ket{\Psi_{ l_{\vec{k}} }} = E_{l_{\vec{k}}} \ket{\Psi_{ l_{\vec{k}} }} \,\qc
  E_{l_{\vec{k}}} = E_0 + \sum_{\vec{s}} l_{\vec{s}} \, \omega_{\vec{s}}
\end{align}

\subsection{Wilson Line expectation values}
Using this notation, one can compute the expectation values of the Wilson line operator $T[Y_n Y_\bn^\dagger]$. 
The Wilson lines originate from the point $x = 0$, which is labeled by the lattice position $n_0$, defined in \cref{n0def}. 
Note that with this notation $x_k$ with $k < n_0$ are negative, while $x_k$ with $k > n_0$ are positive. Note that in this section we assume that we have an odd number of lattice sites, such that we have as many lattice points with $x > 0$ as with $x < 0$. This means that $2n_0 = N-1$. 

As already discussed in the main paper, the Wilson lines on the lattice can be written as
\begin{align}
  \begin{split}
  Y_{n} &= \mathrm{P} \exp\bqty{ ig \var{x}
                   \sum_{i=n_0}^{N-1} \phi_{\vec{n} x_i}(t = x_i - n_0)} \,, \\
  Y_{\bn}^\dagger &= \mathrm{P} \exp\bqty{- ig \var{x}
                             \sum_{i=0}^{n_0} \phi_{\vec{n} x_i}(t = n_0 - x_i)} \,.
  \end{split}
\end{align}

The Wilson line operator is therefore given by
\begin{align}
\label{WilsonOp}
  \begin{split}
  \mathrm{T}[Y_{n} \, Y_{\bn}^\dagger]
    & = e^{ -iH \var{x} \, n_0}
          \exp\bqty{ ig \var{x} \pqty{ \phi_{\vec{n} x_{2n_0}} - \phi_{\vec{n} x_0} }} \\
    & \qquad \times e^{iH \var{x}}
        \exp\bqty{ ig \var{x} \pqty{ \phi_{\vec{n} x_{2n_0-1}} - \phi_{\vec{n} x_1} }} \\
    & \qquad \times \dotsm \times e^{iH \var{x}}
        \exp\bqty{ ig \var{x} \pqty{ \phi_{\var{n} x_{n_0+1}} - \phi_{\var{n} x_{n_0-1}} }} \\
    & \qquad \times e^{iH \var{x}}
        \exp\bqty{ ig \var{x} \pqty{ \phi_{\var{n} x_{n_0}} - \phi_{\var{n} x_{n_0}} }} \\
    & \equiv U_{n_0}(-t) \, U_{n_0-1}(\var{t}) \dotsb U_1(\var{t}) U_0(\var{t}) \,.
  \end{split}
\end{align}
where we have used that $\vec{\bn} = -\vec{n}$ and $x_{n_0+m} = -x_{n_0-m}$.
In the last line we defined
\begin{align}
  U_m(t) \equiv e^{iHt}
    \exp\bqty{ ig \var{x} \pqty{ \phi_{\vec{n}x_{n_0 + m}} - \phi_{\vec{n}x_{n_0 -m}} }} \,,
\end{align}
as well as $t \equiv n_0 \var{x}$, $\var{t} \equiv \var{x}$. 
Note that since $x_{n_0} = 0$ one has $U_{0}(t) = e^{ -i H t}$.

Using \cref{ladderopdef} we can write 
\begin{align}
  \begin{split}
  U_m(t) &= e^{iHt} \prod_{\vec{s}} \exp\!\bqty{
              ig \var{x} \pqty{\frac{\var{p}}{2\pi}}^d
              \pqty{ e^{ i\vec{n} \cdot \vec{p}_{\vec{s}} m \var{x}}
                     - e^{-i \vec{n} \cdot \var{p}_{\vec{s}} m \var{x}}}
              \frac{1}{\sqrt{ 2\omega_{\vec{p}_{\vec{s}}} }}
              \pqty{ a^\dagger_{\vec{p}_{\vec{s}}} - a_{\vec{p}_{\vec{s}}} }} \\
         &= e^{iHt} \prod_{\vec{s}}  \exp\!\bqty{
              -2g \var{x} \pqty{\frac{\var{p}}{2\pi}}^d
              \sqrt{\frac{1}{ 2\omega_{\vec{p}_{\vec{s}}} }}
              \sin(\vec{n} \cdot \vec{p}_{\vec{s}} \, m \var{x})
              \pqty{ a^\dagger_{\vec{p}_{\vec{s}}} - a_{\vec{p}_{\vec{s}}} }} \\
         &= e^{iHt} \prod_{\vec{s}}  D_{\vec{p}_{\vec{s}}} \pqty{\alpha_{\vec{p}_{\vec{s}}}(m)} \,,
  \end{split}
\end{align}
with
\begin{align}
  \alpha_{\vec{p}_{\vec{s}}}(m)
    = 2g \var{x} \pqty{\frac{\var{p}}{2\pi}}^d \sqrt{ \frac{1}{ 2\omega_{\vec{p}_{\vec{s}}} }}
        \sin(\vec{n} \cdot \vec{p}_{\vec{s}} \, m \var{x}) \,.
\end{align}
In the last line we have written the result in terms of the displacement operator that is well know from the theory of coherent states
\begin{align}
  D_{\vec{p}}( \alpha_{\vec{p}} )
    = e^{\alpha_{\vec{p}} a_{\vec{p}}^\dagger - \alpha_{\vec{p}}^* a_{\vec{p}}} \,,
\end{align}
which satisfies the relation
\begin{align}
  D_{\vec{p}}( \alpha_{\vec{p}} ) D_{\vec{p}}( \beta_{\vec{p}} )
    = (2\pi)^d \,D_{\vec{p}}( \alpha_{\vec{p}} + \beta_{\vec{p}} ) \,
        e^{i \Im(\alpha_{\vec{p}} \beta^*_{\vec{p}}) \pqty{\frac{2 \pi}{\var{p}}}^d} \,,
\end{align}
and displacement operators acting on different momentum modes commute
\begin{align}
  [D_{\vec{p}}( \alpha_{\vec{p}} ), D_{\vec{q}}( \beta_{\vec{q}} )] = 0 \,.
\end{align}
The action of the time evolution on the displacement operator is
\begin{align}
  e^{iHt} D_{\vec{p}}[ \alpha_{\vec{p}} ] = D[\alpha_{\vec{p}}(t)] \,\qq{where}
  \alpha_{\vec{p}}(t) = \alpha_{\vec{p}} \, e^{i \omega_{\vec{p}} t} \,.
\end{align}
Combining all this information, one finds
\begin{align}
  \mathrm{T}[Y_n \, Y_{\bn}^\dagger] &= \prod_{\vec{s}} D_{\vec{p}_{\vec{s}}}
    \bqty{\sum_{m=0}^{n_0}\alpha_{\vec{p}_{\vec{s}}}(m) e^{i (n_0-m) \omega_s \var{t}} }
      e^{ i\phi_{\vec{p}_{\vec{s}}} } \,,
\end{align}
with
\begin{align}
  \phi_{\vec{p}_{\vec{s}}} = \pqty{\frac{2\pi}{\var{p}}}^d
    \sum_{n>m} \Im\bqty{ \alpha_{\vec{p}_{\vec{s}}}(n) \, \alpha^*_{\vec{p}_{\vec{s}}}(m) } \,.
\end{align}

To compute expectation values of this operator, we use 
\begin{align}
  \mel{n_{\vec{p}}}{ D_{\vec{p}}[\alpha_{\vec{p}}] }{\ground}
    = \exp\!\bqty{-\pqty{\frac{2\pi}{\var{p}}}^d \frac{\abs{\alpha_{\vec{p}}}^2}{2} }
        \frac{\alpha_{\vec{p}}^{n_p}}{\sqrt{n_{\vec{p}}!}} \,.
\end{align}
In particular, the magnitude square of the vacuum expectation value is given by
\begin{align}
  \begin{split}
\label{finalResultFull}
  \abs{\!\mel{\ground}{ \mathrm{T}[Y_n \, Y_\bn^\dagger }{\ground} }^2
    &= \prod_{\vec{s}} \exp\!\bqty{-\pqty{\frac{2\pi}{\var{p}}}^d
         \abs{ \sum_{m=0}^{n_0}\alpha_{\mathbf{p_s}}(m) e^{i (n_0-m) \omega_{\vec{p}_{\vec{s}}} \var{t}}}^2} \\
    &= \prod_{\vec{s}} \exp\!\bqty{ -4g^2 (\var{x})^2 \pqty{\frac{2\pi}{\var{p}}}^d 
         \frac{1}{2\omega_{\vec{p}_{\vec{s}}}}\left| \sum_{m=0}^{n_0} e^{i (n_0-m) \omega_{\vec{p}_{\vec{s}}} \delta t}  \sin\left(\mathbf{n \cdot p_s} \, m \, \delta x\right)\right|^2} \\
    &= \exp\!\bqty{ -4\frac{g^2}{(2\pi)^d} \sum_{\vec{p}} \frac{1}{2\omega_{\vec{p}}}
         \abs{ \sum_x e^{-i \omega_{\vec{p}} x} \sin(\vec{n}\cdot\vec{p}\, x)}^2} \\
    &= \exp\!\bqty{ -8\frac{g^2}{(2\pi)^d} \sum_{\vec{p}} \frac{1}{2\omega_{\vec{p}}}
         \sum_{x\geq y} \cos(\omega_{\vec{p}}(x-y))
                          \sin(\vec{n}\cdot\vec{p} \, x) \sin(\vec{n}\cdot\vec{p} \, y)} \,,
  \end{split}
\end{align}
where the sum over $x$ and $y$ runs over the lattice position and we have defined 
\begin{align}
  \sum_{x \geq y} f(x,y) \equiv \frac{1}{2} \sum_x f(x,x) + \sum_{x > y} f(x, y) \,.
\end{align}

One can verify that this expression reproduces the perturbative result to order $g^2$. 
To that order, one can write
\begin{align}
  \mel{\ground}{\mathrm{T}[Y_n \, Y_{\bn}^\dagger]}{\ground}
    &= 1 - \frac{g^2}{2} \int_0^L \dd{s} \int_0^L \dd{t}
         \bqty{ D_F(\vec{n} s - \vec{n} t) + D_F(\vec{\bn} s - \vec{\bn} t)
                - D_F(\vec{n} s - \vec{\bn} t)  - D_F(\vec{\bn} s - \vec{n} t) }
\nonumber\\
    & = 1 - g^2 \int_0^L \dd{s} \int_0^s \dd{t}
          \bqty{ D(\vec{n} s - \vec{n} t) + D(\vec{\bn} s - \vec{\bn} t)
                 - D(\vec{n} s - \vec{\bn} t)  - D(\vec{\bn} s - \vec{n} t) }
\nonumber\\
    & = 1 - g^2 \int_0^L \dd{s} \int_0^s \dd{t} \int \frac{\dd[d]{p}}{(2\pi)^d}    
          \frac{1}{2\omega_{\vec{p}}} e^{-i \omega_{\vec{p}} (s-t)}
          \bqty{ e^{i \vec{n} \cdot \vec{p}(s-t)} + e^{-i \vec{n} \cdot \vec{p}(s-t)} 
                 - e^{i \vec{n} \cdot \vec{p}(s+t)} - e^{-i \vec{n} \cdot \vec{p}(s+t)} }
\nonumber\\
    & = 1 - 2g^2 \int_0^L \dd{s} \int_0^s \dd{t} \int \frac{\dd[d]{p}}{(2\pi)^d}   
          \frac{1}{2\omega_{\vec{p}}} e^{-i \omega_{\vec{p}} (s-t)}
          \bqty{ \cos(\vec{n} \cdot \vec{p}(s-t)) -  \cos(\vec{n} \cdot \vec{p}(s+t)) }
\nonumber\\
    & = 1 - 4g^2 \int_0^L \dd{s} \int_0^s \dd{t} \int \frac{\dd[d]{p}}{(2\pi)^d}
          \frac{1}{2\omega_{\vec{p}}} e^{-i \omega_{\vec{p}} (s-t)} 
          \sin(\vec{n} \cdot \vec{p}\,s) \sin(\vec{n} \cdot \vec{p}\, t)
\end{align}
Taking the square of this result one finds
\begin{align}
  \abs{\!\mel{\ground}{\mathrm{T}[Y_n \, Y_{\bn}^\dagger]}{\ground}}^2
    = 1 - 8g^2 \int_0^L \dd{s} \int_0^s \dd{t} \int \frac{\dd[d]{p}}{(2\pi)^d}  
                 \frac{1}{2\omega_{\vec{p}}} \cos(\omega_{\vec{p}} (s-t)) 
                 \sin(\vec{n} \cdot \vec{p}\, s) \sin(\vec{n} \cdot \vec{p}\, t)
\end{align}
which is the continuum limit of \cref{finalResultFull}.

\section{The effect of digitization}
\label{sec:digitization}

In this section we provide some details of the effect of digitization on the Hamiltonian of a free massless field theory.
For this, we consider a (1+1) dimensional field theory with 3 lattice sites at points
\begin{align}
  x_0 = - \var{x}\,\qc
  x_1 = 0 \,\qc
  x_2 = \var{x} \,,
\end{align}
with
Hamiltonian
\begin{align}
  H = \frac{\var{x}}{2} \sum_{i=0}^{2} \bqty{ \dot{\phi}_i^2
                                      - \phi_i \, [\nabla^2 \phi]_j } \,,
\end{align}
The momenta of the 3-site lattice are given by
\begin{align}
  p_0 = -\frac{\pi}{3\var{x}} \,\qc
  p_1 = -\frac{\pi}{\var{x}} \,\qc
  p_2 = \frac{\pi}{\var{x}} \, . 
\end{align}
The corresponding frequencies are obtained from \cref{omega}
\begin{align}
  \omega_0 = \frac{2}{\var{x}} \,\qc
  \omega_1 = \omega_2 = \frac{1}{\var{x}}
  \,.
\end{align}

\begin{figure*}
  \centering
  \includegraphics[width=0.5\linewidth]{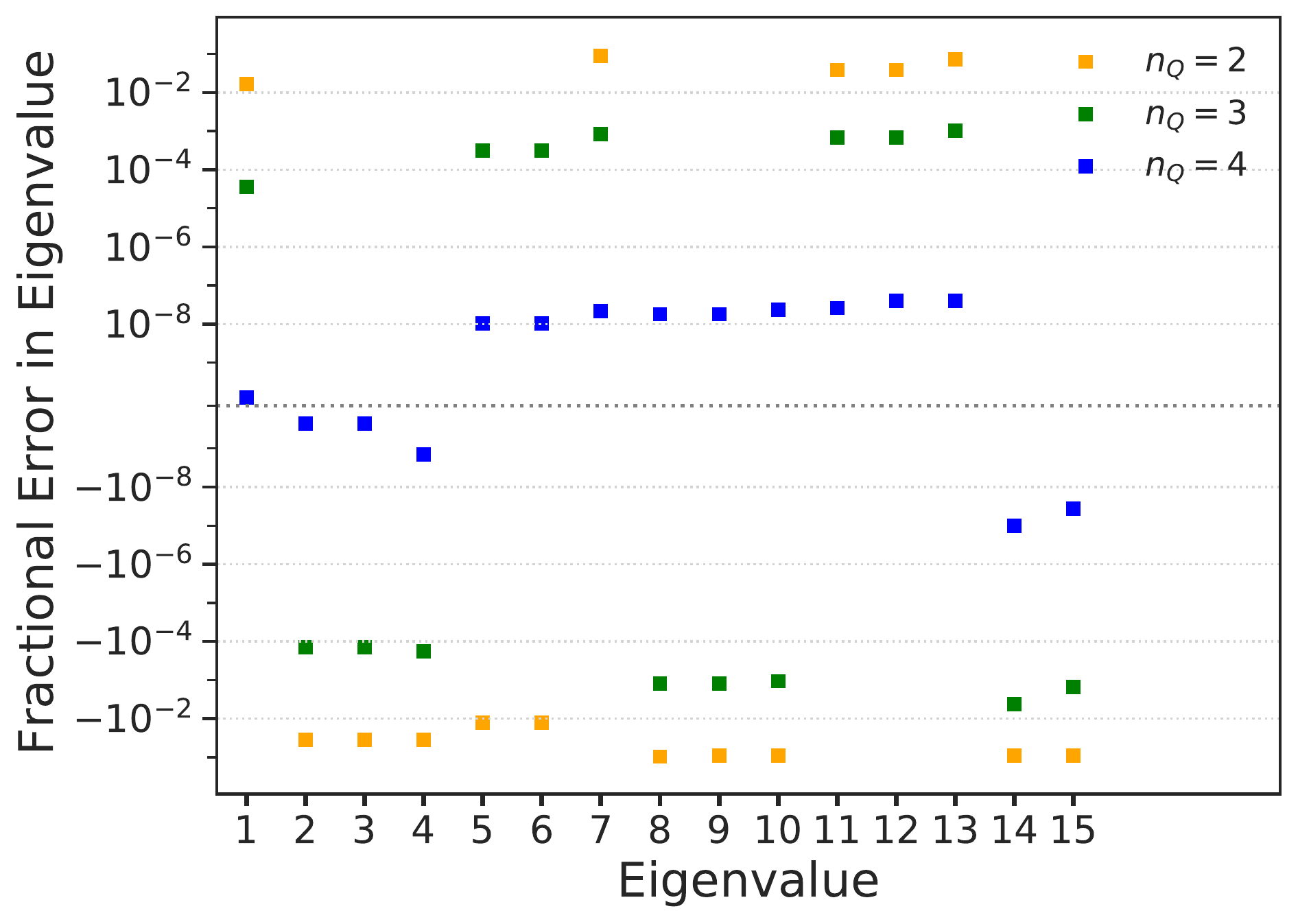}
  \caption{The fractional error in the 15 lowest-lying eigenstates on an $N=3$ lattice. The dependence on the number of qubits per lattice site is shown for $n_Q = 2$ (orange), $n_Q = 3$ (green), and $n_Q = 4$ (blue). The upper (lower) part of the plot displays eigenvalues greater (smaller) than their undigitized values. For $n_Q = 3$, all 2 particles states are already reproduced at better than \SI{1}{\percent} accuracy. (If modes with $\gtrsim \SI{10}{\percent}$ deviations from their continuum values are excluded, this increases to all 5 particle states.) The exponential improvement with number of qubits per site is clearly visible. \label{fig:EigenValues}}
\end{figure*}

The exact eigenvalues of this Hamiltonian are now easily obtained. The energy of the ground state is given by half of the sum of frequencies, and therefore $E_0 = 2 / \var{x}$. The excited states are simply given by the ground state energy increased by up to $n_\phi$ additions of each frequency. Thus, the lowest lying states are
\begin{align}
  1 \times \frac{2}{\var{x}} \,\qc
  2 \times \frac{3}{\var{x}} \,\qc
  4 \times \frac{4}{\var{x}} \,\qc
  6 \times \frac{5}{\var{x}} \,\qc \dotsc
\end{align}
We compare the results obtained for the 15 lowest lying states with the eigenvalues of the above Hamiltonian using $n_q = 2$, $n_q = 3$ and $n_q = 4$ qubits per lattice site in \cref{fig:EigenValues}. One can see that for $n_q = 4$ these lowest states are reproduced with an accuracy of \num{e-7}, for $n_q = 3$ with an accuracy of \num{e-3} and for $n_q = 2$ with an accuracy of about \SI{10}{\percent}. This is in agreement with the findings of~\cite{Klco:2018zqz}.

\begin{figure}
  \centering
  \includegraphics[width=.45\textwidth]{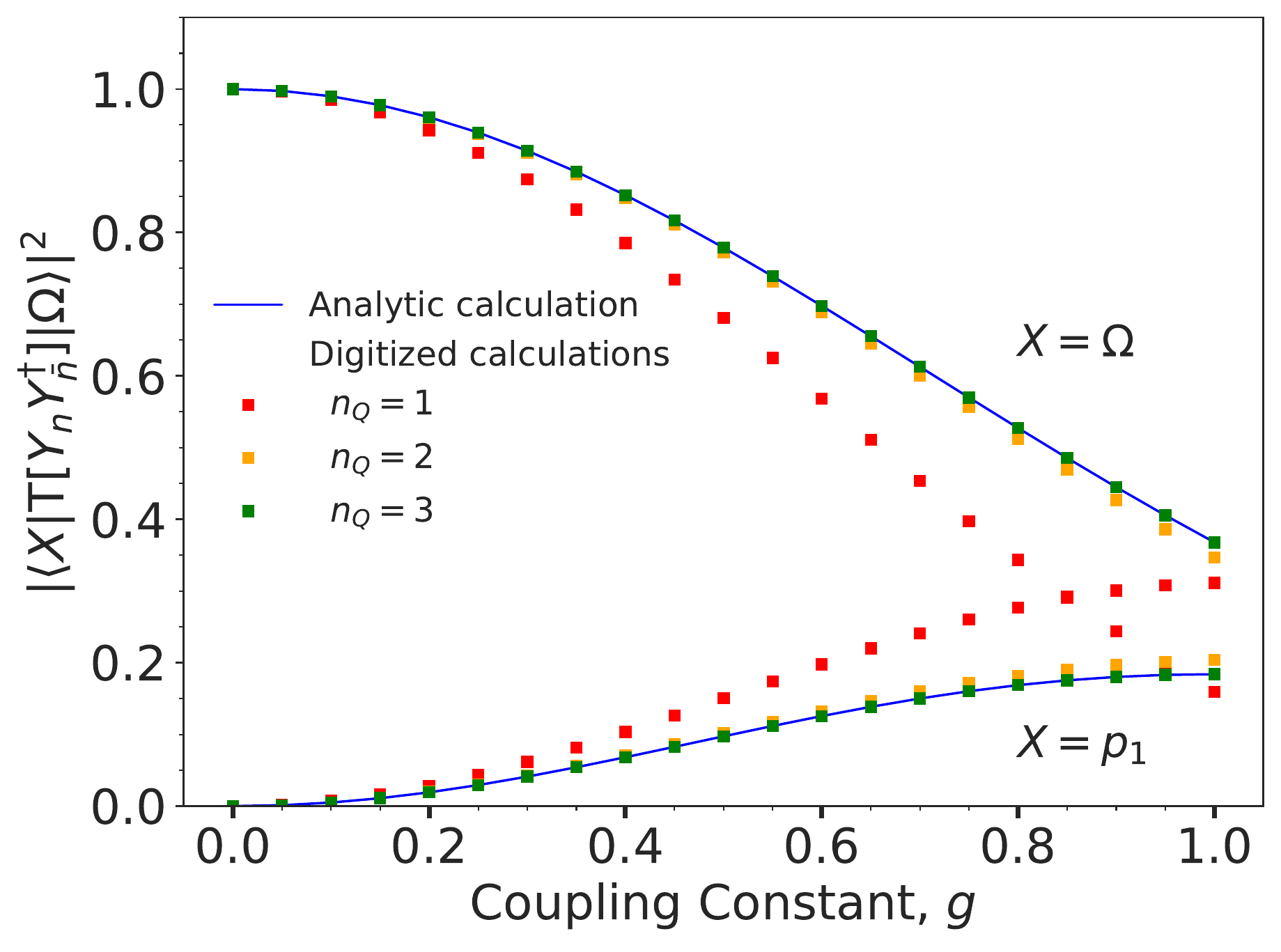}
  \caption{The effect of field digitization on transition rates in the presence of the Wilson line for $N = 3$. The blue solid lines are the analytical result for transitions to the vacuum ($\ground$) and the lowest-lying one one-particle state ($p_1$). Digitization limits the number of states the field can take on for $n_Q = 1$ (red) to 2,$n_Q = 2$ (orange) to 4 and $n_Q = 3$ green to 8. In each case, the exact ground state is digitized and evolved.}
\label{fig:vacExpectSimnQ} 
\end{figure}

\begin{figure}
  \centering
  \includegraphics[width=.45\textwidth]{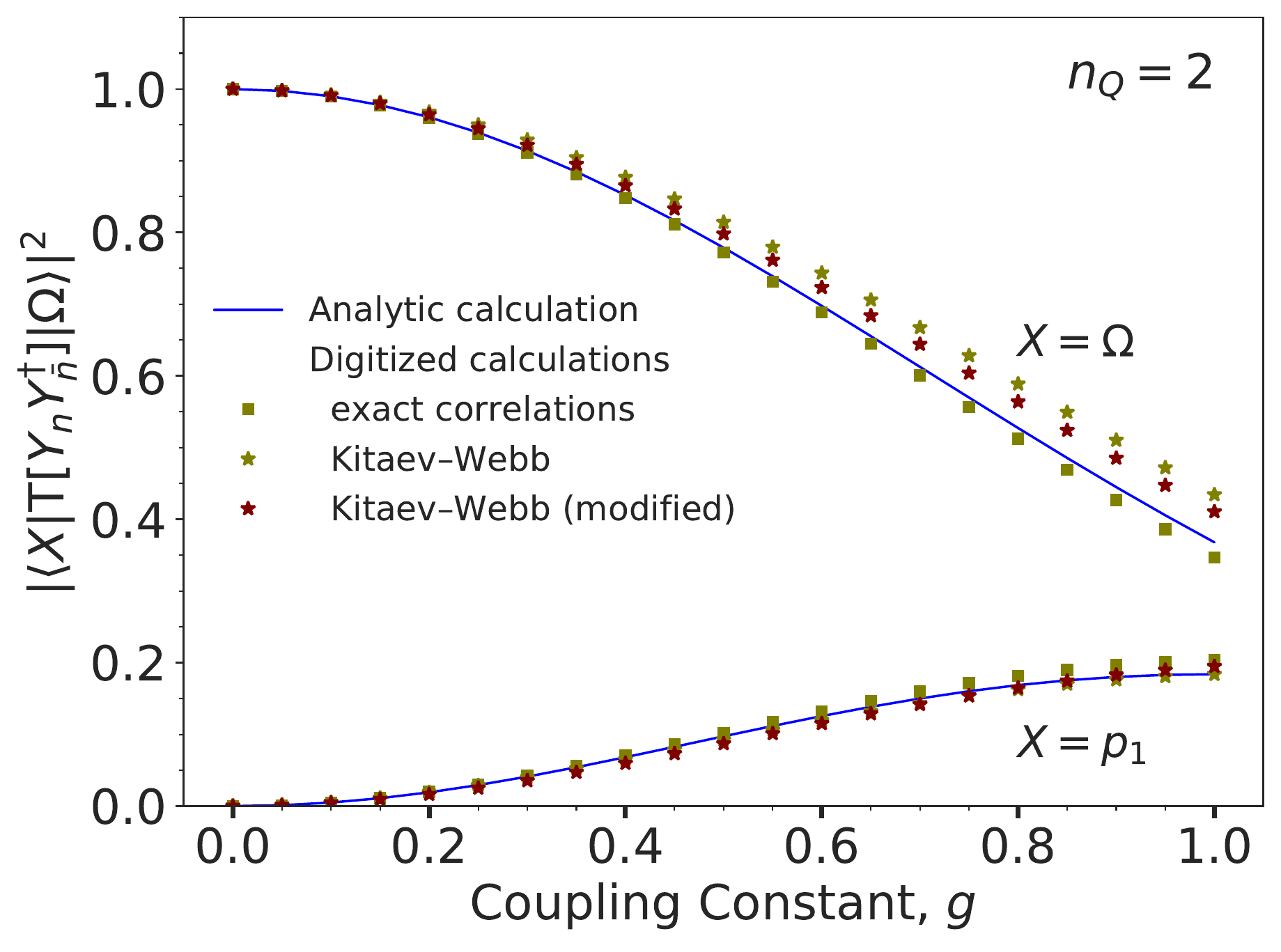}
  \hspace{.05\textwidth}
  \includegraphics[width=.45\textwidth]{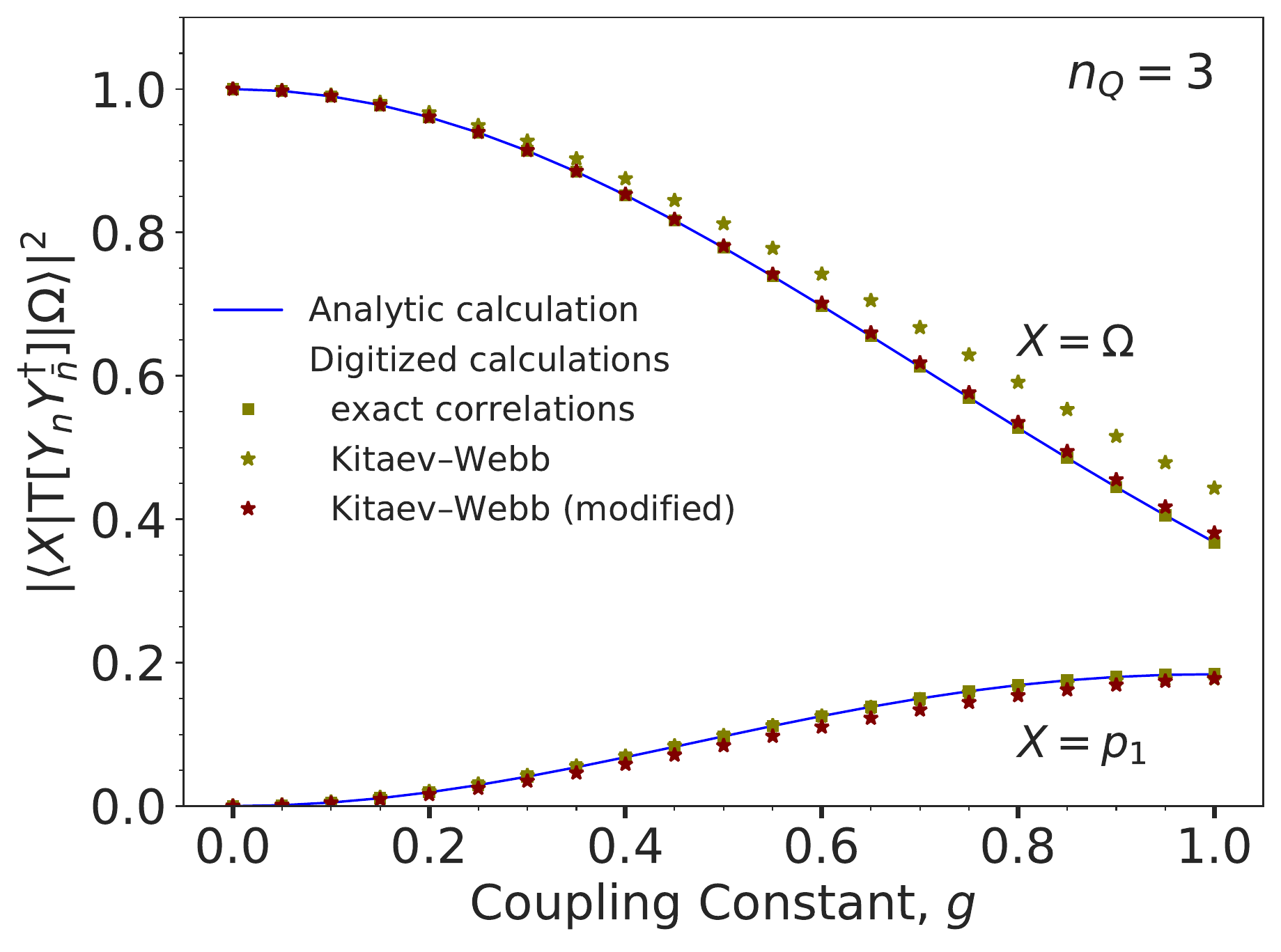}
  \caption{The effect of approximated state preparations on transition rates in the presence of the Wilson line for $N = 3$ lattice sites. The blue solid lines are the analytical result for transitions to the vacuum ($\ground$) and the lowest-lying one one-particle state ($p_1$). We compare the digitization on the exact state with the original and modified KW approximation (see text for details) which can be prepared with polynomial resources in number of qubits. The plots show results for $n_Q = 2$ (left) and $n_Q = 3$ (right).}
\label{fig:vacExpectSimState} 
\end{figure}

In \cref{fig:vacExpectSimnQ} we show the dependence of the vacuum expectation value $\oper_X$ on the coupling $g$ on the number of qubits per lattice site.
One can see that already for $n_Q = 2$ qubits per lattice site one gets a very good approximation to the exact lattice result, and for $n_Q = 3$ the results including digitization are indistinguishable from the exact results if using the exact correlations between sites in the ground state.
\Cref{fig:vacExpectSimState} shows the dependence on the ground state, with the exact ground state shown by the solid squares and the result for the KW approximation shown by stars for $n_Q = 2$ and 3.
One can see that using the KW approximation introduces an additional error, with the resulting emission probability reduced compared to the exact result, but that the resulting difference is sub-percent level by the time $n_Q = 3$.
For these plots only, we compare the KW approximation originally presented in \cite{kitaev2009wavefunction} and used in the rest of the text with a modified form discussed in Section~\ref{subsec:State_prep}.

\section{Consequences of the non-local nature of Wilson lines special considerations in (1+1) dimensions} 
The two Wilson lines $Y_n$ and $Y_\bn^\dagger$ describe how the dynamical soft degrees of freedom in the effective theory (bosons in our case) interact with the fermions that move at the speed of light through the system. 
These Wilson lines contain an integral over the scalar fields along the world lines of the fermions, going from point 0 to $\infty$ along the directions $n^\mu = (1, \vec{n})$ and $\bn^\mu =  (1, -\vec{n})$, and are therefore non-local in nature, with the non-locality extending all the way to infinity.
This non-locality along the light-like directions gives rise to the well known collinear singularities in the matrix elements.
An important consequence of the non-local nature of the operators and their infinite extent is that their lattice implementation necessarily makes them directly sensitive to the lattice volume $L$ and separation $\var{x}$.
The fact that the Wilson lines can only extend up to the edge of the lattice provides a regulator for the angle between the momentum of bosons and the direction \vec{n}, in addition to the IR regulator on the energy of each boson.

As we will now show, this fact gives rise to mixed IR--UV divergences in the lattice definition of the Wilson line operators.
The presence of the finite lattice produces terms $L \, n \cdot p$ and $L \, \bn \cdot p$, regulating divergences as $n \cdot p$ and $\bn \cdot p$ tend to zero.
On the other hand, for finite lattice spacing one has $\omega - \abs{\vec{p}} \sim \abs{\vec{p}}^2 \var{x}^2$, such that the UV regulator is also prohibiting the momenta of the bosons to go on-shell, which in turn also regulates the collinear divergence. 
This implies that collinear divergences in matrix elements give rise to mixed UV-IR divergences, which would be absent in the continuum theory (even though the continuum theory also contains mixed UV-IR divergences coming from the the limit $n \cdot p \to 0$ with $p \to \infty$). 
Once infrared and collinear (IRC) safe physical observables are being calculated, in which all collinear divergences cancel, these lattice artifacts will also cancel, leaving only pure UV divergences related to the normalization of operators.

There is an additional subtlety in a (1+1) dimensional field theory as is used in the numerical results of this paper. 
With only a single spatial dimension one can of course only define a single direction.
This means that in such a theory all momenta have to be aligned with the direction $n$ or $\bn$, which means that every momentum satisfies $n \cdot p = 0$ or $\bn \cdot p = 0$ and therefore gives rise to a collinear divergence.
In other words, each state of the quantum field theory is actually collinear to the directions $n^\mu$ and $\bn^\mu$.
It should be recalled that the underlying short distance process we are describing is the production of two energetic back-to-back particles. 
The differential distribution in the total energy of this process is described by short distance physics captured by the hard function, while the jet function and the soft function discussed in this paper describe the distribution of collinear and soft radiation originating from these energetic particles.
An IRC safe physical observable needs to include all such collinear radiation, which means that in (1+1) dimensions any IRC safe observable has to include all states of the theory, and therefore be completely inclusive of any additional radiation. 
Therefore, the only IRC safe observables are those defined by the hard kinematics, which are completely inclusive over the soft radiation calculated here. Since the Wilson line operator is unitary, the square of the completely inclusive operator is unity and therefore trivial.
In other words, any non-trivial observable in this theory is by construction IRC unsafe, and we therefore choose the simplest such observables, namely the  expectation value of the Wilson line operator between the vacuum states, and the transition from the vacuum to a state with a single momentum mode.

\section{Constructing the quantum circuits}

\subsection{Representation of the Hilbert space and relationship of qubit representation to field values}

To represent the Hilbert space of the massless scalar field theory we discretize the spatial coordinates on a lattice of size $L$ and $N$ sites per dimension. 
The value of the scalar field on each lattice site is digitized and represented by $n_\phi = 2^{n_Q}$ distinct values. 
One can write the field as
\begin{align}
  \phi_i^{(k)} = -\phi_\mathrm{max} + k\, \delta\phi \qfor k = 0, \dotsc , n_\phi  - 1\, \qc 
  \delta\phi = \frac{2\phi_\mathrm{max}}{n_\phi - 1} \,.
\end{align}
Representing the integer $\ket{k}_i$ through its binary representation of the $n_Q$ qubits at lattice site $i$, we can define
\begin{align}
  \hat{\phi}_i \ket{k}_i = \phi_i^{(k)} \ket{k} _i \,.
\end{align}
The value of $\phi_\mathrm{max}$ should be chosen to minimize the minimize the error due to the digitization and depends on the Hamiltonian implemented on the lattice.
The integer state $\ket{k}_i$ is represented as usual through the bitstring of the $n_Q$ qubits at the given lattice site
\begin{align}
  \ket{k}_i = \ket{q_0}_i \dotsb \ket{q_{n_Q-1}}_i
  \,.
\end{align}

The full Hilbert space is then represented through the tensor product of the states $\ket{k}$ on each lattice site
\begin{align}
  \ket{\Psi} = \ket{k}_0 \dotsb \ket{k}_{N^d}
  \,.
\end{align}
Our explicit circuit constructions in this paper will only use a single spatial direction, such that we use
\begin{align}
  \ket{\Psi} = \ket{k}_0 \dotsb \ket{k}_{N}
  \,.
\end{align}
%

\subsection{The free Hamiltonian}
The construction of the Hamiltonian of free massless scalar field theory follows previous work~\cite{Macridin:2018gdw,Macridin:2018oli,Klco:2018zqz}. 
One can easily convince oneself that the operator $\hat{\phi}_i$ can be written through its action on the $n_Q$ qubits at each lattice site
\begin{align}
\label{phidef}
  \hat \phi_i = \sum_{j=0}^{n_Q - 1} 2^j \hat \sigma_{z,i}^{(j)} \,,
\end{align}
where the operator $\hat{\sigma}_{z,i}^{(j)}$ is a single $\sigma_z$ Pauli matrix applied to the $j$th qubit of the $i$th lattice register.

The Hamiltonian is a sum over two pieces that do not commute with one another. The time derivative of the field is the conjugate field $\pi_i \equiv \dot{\phi}_i$, and one can write
\begin{align}
  H = H_\pi + H_\phi
\end{align}
with
\begin{align}
  H_\pi = \var{x} \sum_i \pi_i^2 \,\qc
  H_\phi = \var{x} \sum_{i=0}^{N-1} \phi_i [\nabla^2 \phi]_i  \,.
\end{align}
As discussed before, the $\phi_i [\nabla^2 \phi]_i$ operator only requires nearest neighbor interactions on the lattice. 
The time evolution is then written in terms of the Suzuki--Trotter formula (we give the first order expression here)
\begin{align}
\label{eq:HTrotter}
  \bqty{ e^{-iHt} }_n = \bqty{ e^{i H_\pi t/n} \, e^{i H_\phi t/n} }^n \,.
\end{align}

To construct the exponential of the $H_\pi$ operator, we use that the $\phi$ and $\pi$ are related by a Fourier transform on the $\phi$ register. Thus, we can write
\begin{align}
  e^{i H_\pi t} = \mathrm{QFT}^{-1} \, e^{i \var{x}\, t \phi_i^2 } \, \mathrm{QFT} \,,
\end{align}
where $\mathrm{QFT}$ denotes the (symmetrized) Quantum Fourier Transform, which was discussed in~\cite{Klco:2018zqz}. We do not repeat its circuit here. 

Given this, on needs to find a circuit representation $\exp[i\theta \phi_i \phi_j]$ for general $i$ and $j$, from which one can construct both the circuits for $\exp[iH_\pi t]$ and $\exp[iH_\phi t]$. 

Using \cref{phidef}, one can write
\begin{align}
  \hat{\phi}_i \hat{\phi}_j
    &= \bqty{ \sum_{l=0}^{n_Q - 1} 2^l \hat \sigma_{z,i}^{(l)} }
         \bqty{ \sum_{k=0}^{n_Q - 1} 2^l \hat \sigma_{z,j}^{(k)} }
     = \sum_{l=0}^{n_Q - 1} \sum_{k=0}^{n_Q - 1} 2^{(l+k)} \sigma_{z,i}^{(l)} \sigma_{z,j}^{(k)} \,.
\end{align}
This allows us to write
\begin{align}
\label{expphi2}
  \exp[i \theta \hat \phi_i \hat \phi_j]
    &= \prod_{l=0}^{n_Q - 1}\prod_{k=0}^{n_Q - 1}
         \exp\!\bqty{ i\, 2^{(l+k)}\theta\, \sigma_{z,i}^{(l)}\sigma_{z,j}^{(k)} } \,.
\end{align}
The action $\exp\!\bqty{ i\, 2^{(k+l)}\theta\, \sigma_{z,i}^{(l)}\sigma_{z,j}^{(k)} }$ is equal to $\exp[ i 2^{(k+l)}\theta ]$ if qubits $q_i^{(l)}$ and $q_i^{(k)}$ are equal and $\exp[ -i 2^{(k+l)}\theta ]$ if they are opposite. Thus, it can be implemented by the circuit
\begin{align*}
  \Qcircuit @C=0.5em @R=0.8em @!R{
  \lstick{\ket{l}_i} & \ctrl{1} & \qw                                  & \ctrl{1} & \qw \\
  \lstick{\ket{k}_i} & \targ    & \gate{e^{-i \pqty{2^{k+l}\theta} Z}} & \targ    & \qw \\
  }
\end{align*}
and the full operator $\exp\!\bqty*{i\theta \hat{\phi}_i \hat{\phi}_j}$ for $n_q$ qubits per lattice site is therefore implemented by stringing together the $n_Q(n_Q-1)/2$ different possible 2-qubit circuits shown above.

\subsection{Ground state preparation}
\label{subsec:State_prep}

The ground state of a massless scalar field theory is given by a multivariate Gaussian 
\begin{align}
  \ket{\Psi} = \exp\bqty{ -\frac{1}{2} \hat{\phi}_i G_{ij} \hat{\phi}_j }
                 \ket{k_0} \dotsb \ket{k}_n
\end{align}
While \textsc{Qiskit} provides a function to generate an aribtrary Gaussian multivariate distribution, the number of gates in the resulting circuit scale exponentially with the number of qubits in the system. 
However, an algorithm with polynomial scaling was derived by Kitaev and Webb~\cite{kitaev2009wavefunction}. %
While this algorithm does not produce the exact multivariate distribution in its digitized form, it approaches the correct limit as the number of qubits per lattice site becomes large.

The KW algorithm relies on the LDL or square-root-free Cholesky decomposition, which rewrites the correlation matrix in terms of a diagonal matrix $D$ and a lower unit-triangular matrix $L$ (a lower triangular matrix with 1's on the diagonal)
\begin{align}
  G = L D L^\dagger
\end{align}
An arbitrary multi-dimensional Gaussian distribution can then be created by generating a series of uncorrelated Gaussian distributions according to the diagonal matrix $D$, and then applying shear matrices $L$ through a remapping of the basis states of the Hilbert space. 
For details of this algorithm, we refer the reader to the original paper. 
In Section~\ref{sec:digitization} we mentioned a modified version of the KW procedure, which only rounds the shear matrices applied to the uncorrelated states to the nearest digitized field value after applying the the full shear matrix rather than after every individual shearing operation, as was originally proposed in~\cite{kitaev2009wavefunction}. 
While requiring more ancilla qubits for memory, this does not affect the polynomial scaling of the approximation and results in exponentially greater fidelity with the exact ground state for the $n_Q \ge 3$ cases~\cite{futureKW}.

For 2 qubits per lattice site, the second step of the KW algorithm does not actually change the basis states, such that KW state preparation is equal to the production of a uncorrelated Gaussians at each lattice site with width given by the diagonal entries of the matrix $D$. In this work, we use the Cholesky decomposition of the correlation matrix, but rather than using the full KW algorithm to produce the uncorrelated Gaussian, we use the built in functionality of \textsc{Qiskit}, even though it has exponential scaling with $n_Q$. 

\subsection{Wilson line operator}
The Wilson line operator on a lattice was given in \cref{WilsonOp}. 
From this expression on can see that it is determined through the successive application of the operator $\exp[ ig \var{x} \phi_n]$ and Hamiltonian evolution. 
Given the circuit for Hamiltonian evolution derived above, one therefore only needs the circuit for the exponential of a single field.
This is easily derived using a similar derivation to the Hamiltonian case, and one writes
\begin{align}
\label{expphi}
  \exp\!\bqty*{i\theta \hat{\phi}_i}
    = \prod_{j=0}^{n_Q - 1} \exp\!\bqty{ i 2^{j} \theta \sigma_{z,i}^{(j)}} \,, 
\end{align}
which can easily be written in circuit form
\begin{align*}
  \Qcircuit @C=0.5em @R=0.8em @!R{
  \lstick{\ket{0}_i}     & \gate{e^{-i\theta Z}}              & \qw    \\
  \vdots                 & \dotsb                             & \vdots \\
  \lstick{\ket{n_Q-1}_i} & \gate{e^{-i 2^{(n_q-1)} \theta Z}} &  \qw
}
\end{align*}

\section{Validation of the circuits for the Hamiltonian}
\begin{figure*}
  \centering
  \includegraphics[width=0.45\linewidth]{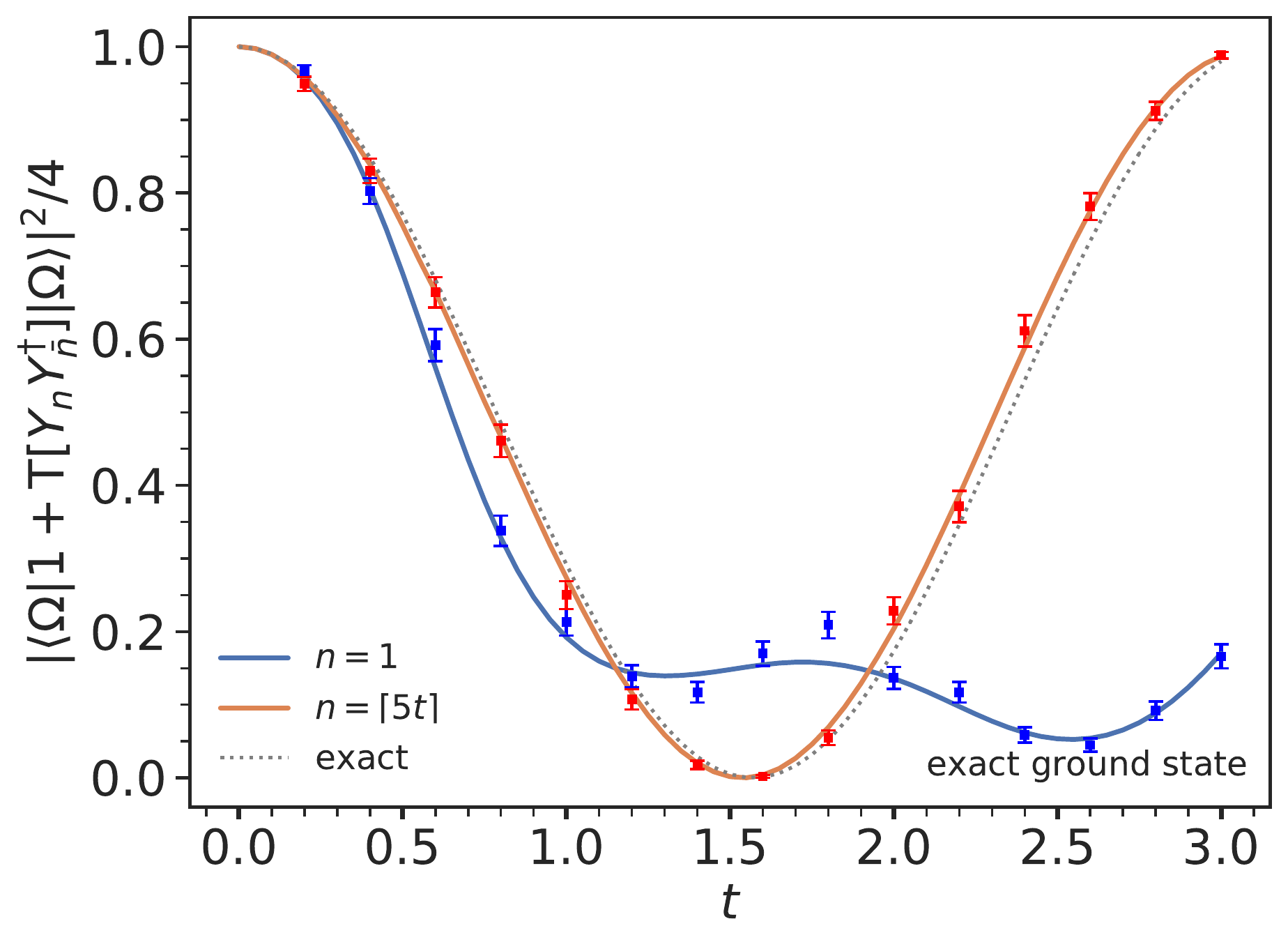}
  \includegraphics[width=0.45\linewidth]{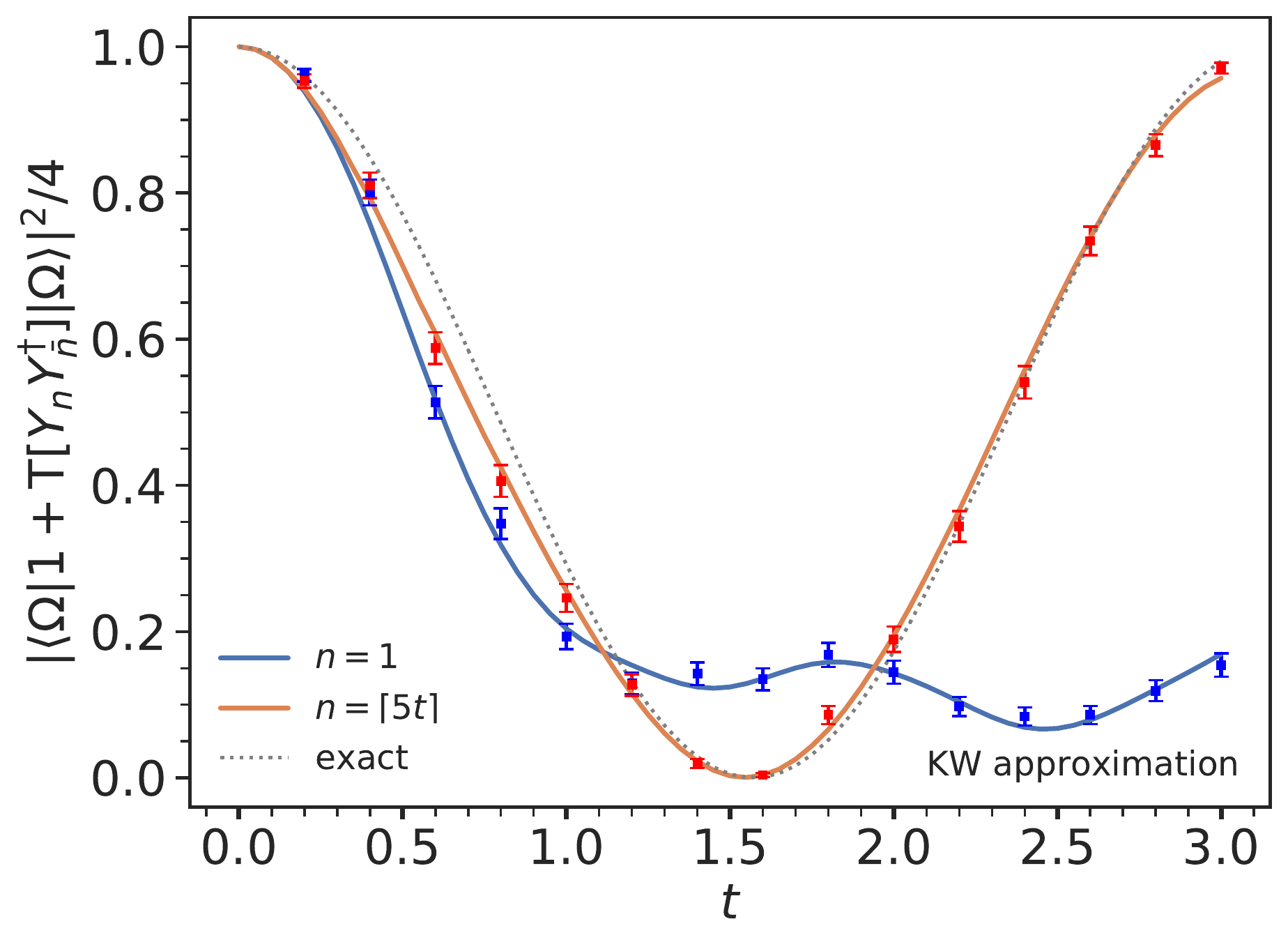}
  \caption{The value of $\abs{1 + \mel{\ground}{\bqty{e^{-iHt}}_n }{\ground}}^2/4$ for two different values of $n$. The blue line shows the result for $n = 1$, while the red line shows the result for $n = \ceil{t / 0.2}$. Measurements from a noiseless simulation of 512 shots/point are overlayed. On the left, we show the result for the exact ground state, while on the right we show the result for the Kitaev--Webb approximation. In dashed black we show the exact result.
\label{figGroundStateOverlap}
}
\end{figure*}

As a validation of the quantum circuits we first check the implementation of the free field theory (ground state preparation and time evolution). 
In particular, we check to what degree the ground state is an eigenstate of the Hamiltonian, and how well the energy of the ground state agrees with the known analytical value. 
We begin by computing the overlap
\begin{align}
  f(t) = \abs{\!\mel{\ground}{\bqty{e^{-iHt}}_n }{\ground}}^2 \,,
\end{align}
where $\bqty{e^{-i H t}}_n$ is the Trotterized Hamiltonian with $n$ steps given in \cref{eq:HTrotter}. 
This is implemented with the circuit
\begin{align*}
  \Qcircuit @C=0.5em @R=0.8em @!R{
    & \qw {/} & \gate{U_\mathrm{state}} & \qw & \gate{\bqty{e^{-iHt}}_n}
    & \qw & \gate{U_\mathrm{state}^\dagger} & \qw  & \meter
}
\end{align*}
and the function $f(t)$ is obtained by the fraction of measurements where all qubits are back the initial $\ket{0}$ state.
If the ground state is indeed an eigenstate of the Hamiltonian, and the Trotterized Hamiltonian is equal to the full Hamiltonian, this function should be identically 1, and any deviation from this value should be due to Trotterization errors or because the state $\ket{\ground}$ not being equal to the true ground state.
In \cref{figGroundStateOverlap} we show this overlap for the exact ground state on the left and for the KW ground state on the right.
The result on the left confirms that with more Trotter steps we approach unity, while the result on the right shows that even for a large amount of Trotter steps the KW approximation leads to deviations from unity.
\begin{figure*}
  \centering
  \includegraphics[width=0.45\linewidth]{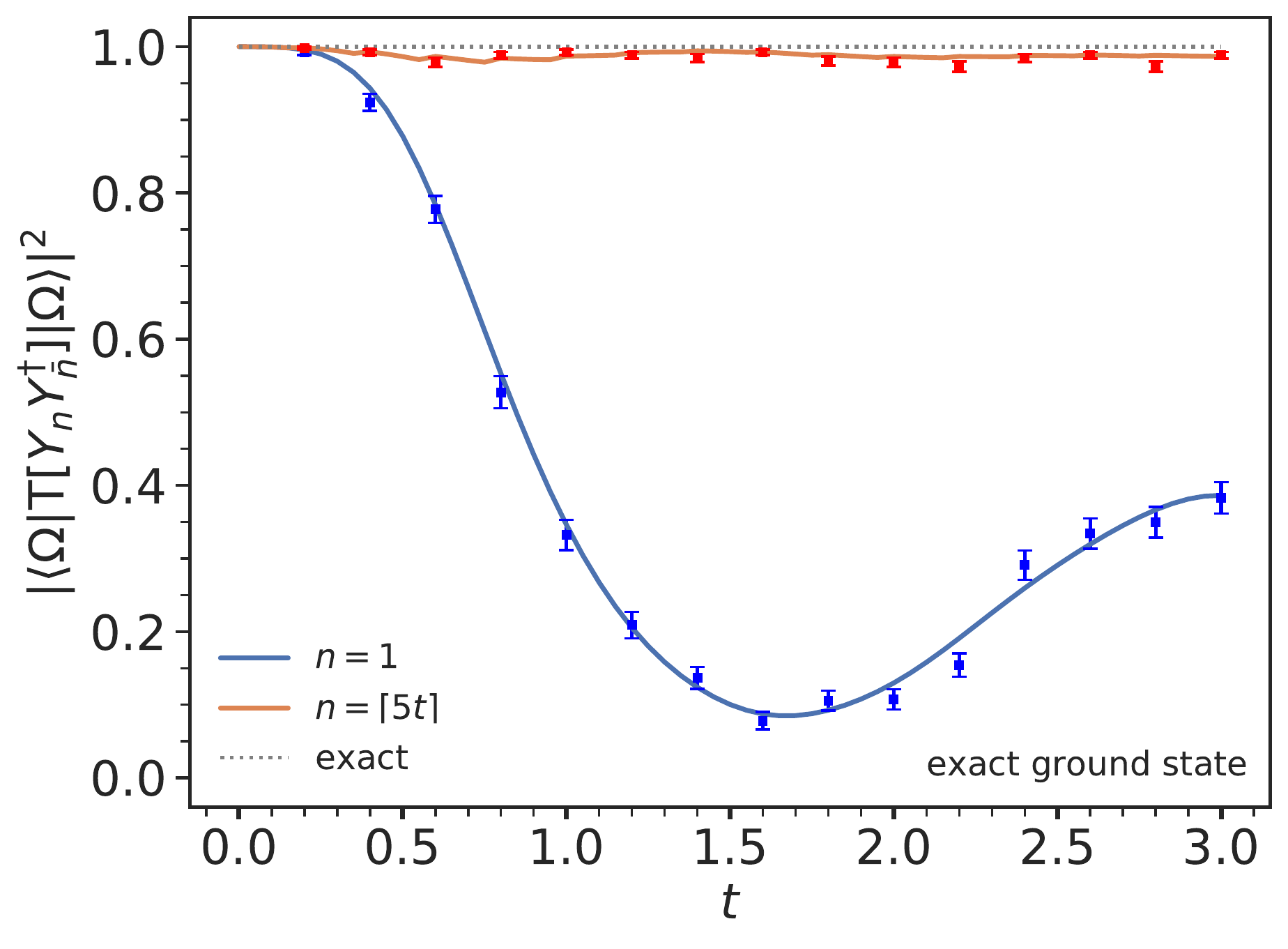}
  \includegraphics[width=0.45\linewidth]{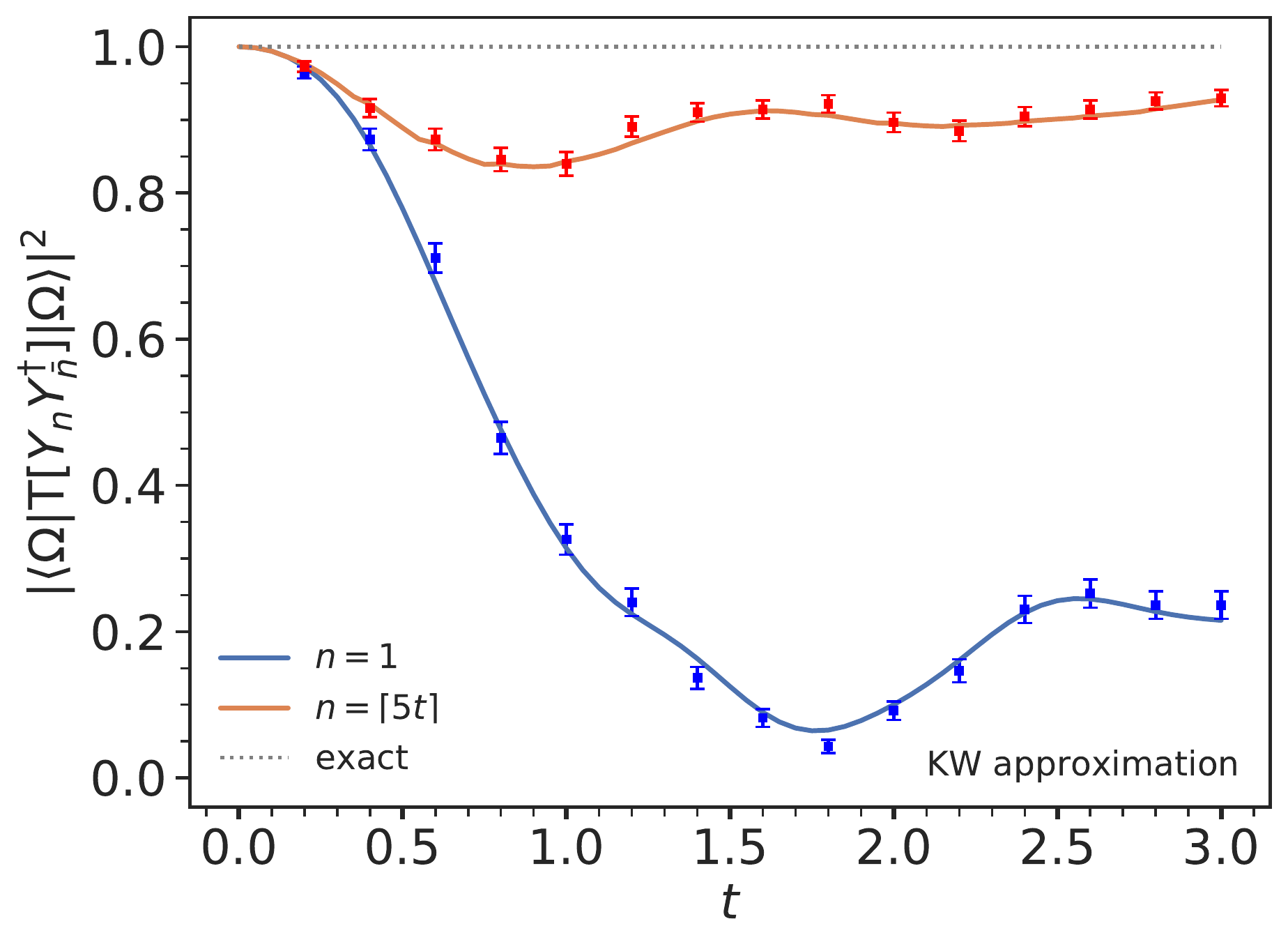}
  \caption{The value of $\abs{\!\mel{\ground}{\bqty{e^{-iHt}}_n }{\ground}}^2$ for two different values of $n$. The blue line shows the result for $n = 1$, while the red line shows the result for $n = \ceil{t / 0.2}$. Measurements from a noiseless simulation of 512 shots/point are overlayed. On the left, we show the result for the exact ground state, while on the right we show the result for the KW approximation. For the exact ground state, the deviation from unity is only due to the Trotterization of the Hamiltonian, and gets smaller with more Trotter steps (and more quantum gates). Since the KW approximation is not an eigenstate of the full Hamiltonian, the deviation from unity is due to both this approximation and the Trotterization.
\label{figGroundStateOverlap2}
}
\end{figure*}

While this measurements tests to what degree the ground state is an eigenstate of the Trotterized Hamiltonian, it can not check for the energy of the ground state since that manifests itself only as a pure phase in the above circuit. A slight variation of this circuit 
\begin{align*}
  \Qcircuit @C=0.5em @R=0.8em @!R{
    & \qw {/} &  \gate{U_\mathrm{state}} & \qw & \gate{\bqty{e^{-iHt}}_n} &  \qw
    & \gate{U_\mathrm{state}^\dagger} &  \qw  & \meter \\
    & \qw     &  \gate{H}                & \qw & \ctrl{-1}                &  \qw
    & \gate{H}                        &  \qw     & \meter
  }
\end{align*}
can measure the energy of the ground state. The fraction of measurements with all qubits in the $\ket{0}$ state is given by
\begin{align}
  f_\mathrm{ctr}(t) = \frac{1}{4} \abs{1 + \mel{\ground}{\bqty{e^{-iHt}}_n }{\ground}}^2 \,,
\end{align}
which is sensitive to the energy $E_\ground$. One can see that using the digitization of exact ground state and sufficient Trotter steps one produces the analytically expected dependence on the ground state energy up to the small difference in period due to the shift in ground state energy due to digitization. Conversely, with the KW states one also sees a small reduction in probability due to leakage out of the approximate ground state, as expected.

\section{Error Mitigation on IBMQ}

We mitigate both readout errors and gate errors.  
Readout error mitigation proceeds with a classical post-processing step.  
We prepare all $2^{n_\text{qubit}}$ possible states $\ket{i}$ and measure the frequency of observing the state $\ket{f}$.   
These probabilities are tabulated in a $2^{n_\text{qubit}}\times 2^{n_\text{qubit}}$ matrix called the response matrix $R$.  
The observations from the main experiment are represented as a vector $m_i$ and then readout error mitigation proceeds iteratively~\cite{ReadoutCorrection,DAgostini:1994fjx,1974AJ.....79..745L,Richardson:72}:
\begin{align}
\label{eq:IBU}
    t_i^{n+1}&=\sum_j\Pr(\text{truth is $i$} | \text{measure $j$}) \times m_j\nonumber\\
    &=\sum_j\frac{R_{ji}t_i^n}{\sum_k R_{jk}t_k^n}\times m_j\,,
\end{align}
where $t_i^0=1/2^{n_\text{qubit}}$ is the uniform prior and the number of iterations $n$ is chosen to be 20. 
The result is not sensitive to small variations in these choices.  
The iterative procedure in Eq.~\ref{eq:IBU} avoids pathologies from other forms of regularized matrix inversion and can be further improved with variations like readout rebalancing~\cite{2010.07496} and subexponential approximations~\cite{geller_efficient_2020,song_10-qubit_2017,gong_genuine_2019,wei_verifying_2020,hamilton2020scalable}.
Matrix inversion approaches from the quantum simulators \texttt{pyQuil}~\cite{pyquil} (by Rigetti), \texttt{Cirq}~\cite{cirq,arute2020quantum} (by Google), and \texttt{XACC}~\cite{alex2019xacc,xacc,mccaskey_quantum_2019} and the least squares method from \texttt{Qiskit} by IBM~\cite{Qiskit,ignis} (see also Ref.~\cite{1904.11935,1907.08518}) produce similar results~\cite{ReadoutCorrection}.
In current IBMQ machines, the dominant gate noise is from multiqubit operations, which are only the CNOT gates in our experiments.  
A common model for CNOT errors is the quantum depolarizing noise channel.  
This error is mitigated by systematically increasing the noise and then extrapolating to zero noise.  
In particular, we remove the leading order depolarizing noise by repeating the nominal measurement with a circuit where every CNOT gate is replaced with 3 CNOT gates.  
Following Ref.~\cite{He:2020udd}, we then set the mitigated result as $3/2$ times the nominal circuit minus $1/2$ the result from the circuit result with triple the CNOT gates.  
This fixed identity insertion method (FIIM) was first proposed in Ref.~\cite{Dumitrescu:2018} and multiple variations exist that improve the fidelity of the extrapolation or are more resource efficient~\cite{PhysRevX.8.031027,PhysRevLett.119.180509,He:2020udd}.  


\begin{thebibliography}{33}%
\makeatletter
\providecommand \@ifxundefined [1]{%
 \@ifx{#1\undefined}
}%
\providecommand \@ifnum [1]{%
 \ifnum #1\expandafter \@firstoftwo
 \else \expandafter \@secondoftwo
 \fi
}%
\providecommand \@ifx [1]{%
 \ifx #1\expandafter \@firstoftwo
 \else \expandafter \@secondoftwo
 \fi
}%
\providecommand \natexlab [1]{#1}%
\providecommand \enquote  [1]{``#1''}%
\providecommand \bibnamefont  [1]{#1}%
\providecommand \bibfnamefont [1]{#1}%
\providecommand \citenamefont [1]{#1}%
\providecommand \href@noop [0]{\@secondoftwo}%
\providecommand \href [0]{\begingroup \@sanitize@url \@href}%
\providecommand \@href[1]{\@@startlink{#1}\@@href}%
\providecommand \@@href[1]{\endgroup#1\@@endlink}%
\providecommand \@sanitize@url [0]{\catcode `\\12\catcode `\$12\catcode
  `\&12\catcode `\#12\catcode `\^12\catcode `\_12\catcode `\%12\relax}%
\providecommand \@@startlink[1]{}%
\providecommand \@@endlink[0]{}%
\providecommand \url  [0]{\begingroup\@sanitize@url \@url }%
\providecommand \@url [1]{\endgroup\@href {#1}{\urlprefix }}%
\providecommand \urlprefix  [0]{URL }%
\providecommand \Eprint [0]{\href }%
\providecommand \doibase [0]{http://dx.doi.org/}%
\providecommand \selectlanguage [0]{\@gobble}%
\providecommand \bibinfo  [0]{\@secondoftwo}%
\providecommand \bibfield  [0]{\@secondoftwo}%
\providecommand \translation [1]{[#1]}%
\providecommand \BibitemOpen [0]{}%
\providecommand \bibitemStop [0]{}%
\providecommand \bibitemNoStop [0]{.\EOS\space}%
\providecommand \EOS [0]{\spacefactor3000\relax}%
\providecommand \BibitemShut  [1]{\csname bibitem#1\endcsname}%
\let\auto@bib@innerbib\@empty

\bibitem{Jordan:2011ne}
S.~P.~Jordan, K.~S.~M.~Lee and J.~Preskill,
Science \textbf{336}, 1130-1133 (2012)
[arXiv:1111.3633 [quant-ph]].

\bibitem{Kogut:1974ag}
J.~B.~Kogut and L.~Susskind,
Phys. Rev. D \textbf{11}, 395-408 (1975)

\bibitem{Kogut:1979wt}
J.~B.~Kogut,
Rev. Mod. Phys. \textbf{51}, 659 (1979)

\bibitem{Bronzan:1984xb}
J.~B.~Bronzan,
Phys. Rev. D \textbf{31}, 2020-2028 (1985)

\bibitem{Jordan:2011ci}
S.~P.~Jordan, K.~S.~M.~Lee and J.~Preskill,
Quant. Inf. Comput. \textbf{14}, 1014-1080 (2014)
[arXiv:1112.4833 [hep-th]].

\bibitem{Macridin:2018gdw}
A.~Macridin, P.~Spentzouris, J.~Amundson and R.~Harnik,
Phys. Rev. Lett. \textbf{121}, no.11, 110504 (2018)
[arXiv:1802.07347 [quant-ph]].

\bibitem{Macridin:2018oli}
A.~Macridin, P.~Spentzouris, J.~Amundson and R.~Harnik,
Phys. Rev. A \textbf{98}, no.4, 042312 (2018)
[arXiv:1805.09928 [quant-ph]].

\bibitem{Klco:2018zqz}
N.~Klco and M.~J.~Savage,
Phys. Rev. A \textbf{99}, no.5, 052335 (2019)
[arXiv:1808.10378 [quant-ph]].

\bibitem{Hackett:2018cel}
D.~C.~Hackett, K.~Howe, C.~Hughes, W.~Jay, E.~T.~Neil and J.~N.~Simone,
Phys. Rev. A \textbf{99}, no.6, 062341 (2019)
[arXiv:1811.03629 [quant-ph]].

\bibitem{Yeter-Aydeniz:2018mix}
K.~Yeter-Aydeniz, E.~F.~Dumitrescu, A.~J.~McCaskey, R.~S.~Bennink, R.~C.~Pooser and G.~Siopsis,
Phys. Rev. A \textbf{99}, no.3, 032306 (2019)
[arXiv:1811.12332 [quant-ph]].

\bibitem{Kreshchuk:2020dla}
M.~Kreshchuk, W.~M.~Kirby, G.~Goldstein, H.~Beauchemin and P.~J.~Love,
[arXiv:2002.04016 [quant-ph]].

\bibitem{Kreshchuk:2020kcz}
M.~Kreshchuk, S.~Jia, W.~M.~Kirby, G.~Goldstein, J.~P.~Vary and P.~J.~Love,
[arXiv:2009.07885 [quant-ph]].

\bibitem{Haase:2020kaj}
J.~F.~Haase, L.~Dellantonio, A.~Celi, D.~Paulson, A.~Kan, K.~Jansen and C.~A.~Muschik,
[arXiv:2006.14160 [quant-ph]].

\bibitem{Gross:1973id}
D.~J.~Gross and F.~Wilczek,
Phys. Rev. Lett. \textbf{30}, 1343-1346 (1973)

\bibitem{Politzer:1973fx}
H.~D.~Politzer,
Phys. Rev. Lett. \textbf{30}, 1346-1349 (1973)

\bibitem{Collins:1984kg}
J.~C.~Collins, D.~E.~Soper and G.~F.~Sterman,
Nucl. Phys. B \textbf{250}, 199-224 (1985)

\bibitem{Catani:1992ua}
S.~Catani, L.~Trentadue, G.~Turnock and B.~R.~Webber,
Nucl. Phys. B \textbf{407}, 3-42 (1993)

\bibitem{Bonciani:2003nt}
R.~Bonciani, S.~Catani, M.~L.~Mangano and P.~Nason,
Phys. Lett. B \textbf{575}, 268-278 (2003)
[arXiv:hep-ph/0307035 [hep-ph]].

\bibitem{Marchesini:1983bm}
G.~Marchesini and B.~R.~Webber,
Nucl. Phys. B \textbf{238}, 1-29 (1984)

\bibitem{Bengtsson:1986et}
M.~Bengtsson and T.~Sjostrand,
Nucl. Phys. B \textbf{289}, 810-846 (1987)

\bibitem{Bauer:2019qxa}
C.~W.~Bauer, W.~A.~De Jong, B.~Nachman and D.~Provasoli,
[arXiv:1904.03196 [hep-ph]].

\bibitem{Dumitrescu:2018njn}
E.~F.~Dumitrescu, A.~J.~McCaskey, G.~Hagen, G.~R.~Jansen, T.~D.~Morris, T.~Papenbrock, R.~C.~Pooser, D.~J.~Dean and P.~Lougovski,
Phys. Rev. Lett. \textbf{120}, no.21, 210501 (2018)
[arXiv:1801.03897 [quant-ph]].


\bibitem{Lu:2018pjk}
H.~H.~Lu, N.~Klco, J.~M.~Lukens, T.~D.~Morris, A.~Bansal, A.~Ekstr\"om, G.~Hagen, T.~Papenbrock, A.~M.~Weiner and M.~J.~Savage, \textit{et al.}
Phys. Rev. A \textbf{100}, no.1, 012320 (2019)
doi:10.1103/PhysRevA.100.012320
[arXiv:1810.03959 [quant-ph]].

\bibitem{Cervia:2019res}
M.~J.~Cervia, A.~V.~Patwardhan, A.~B.~Balantekin, S.~N.~Coppersmith and C.~W.~Johnson,
Phys. Rev. D \textbf{100}, no.8, 083001 (2019)
[arXiv:1908.03511 [hep-ph]].

\bibitem{Roggero:2019myu}
A.~Roggero, A.~C.~Y.~Li, J.~Carlson, R.~Gupta and G.~N.~Perdue,
Phys. Rev. D \textbf{101}, no.7, 074038 (2020)
[arXiv:1911.06368 [quant-ph]].

\bibitem{Cervia:2020fkk}
M.~J.~Cervia, A.~B.~Balantekin, S.~N.~Coppersmith, C.~W.~Johnson, P.~J.~Love, C.~Poole, K.~Robbins and M.~Saffman,
[arXiv:2011.04097 [hep-th]].

\bibitem{Stetina:2020abi}
T.~F.~Stetina, A.~Ciavarella, X.~Li and N.~Wiebe,
[arXiv:2101.00111 [quant-ph]].

\bibitem{Bauer:2000ew}
C.~W.~Bauer, S.~Fleming and M.~E.~Luke,
Phys. Rev. D \textbf{63}, 014006 (2000)
[arXiv:hep-ph/0005275 [hep-ph]].

\bibitem{Bauer:2000yr}
C.~W.~Bauer, S.~Fleming, D.~Pirjol and I.~W.~Stewart,
Phys. Rev. D \textbf{63}, 114020 (2001)
[arXiv:hep-ph/0011336 [hep-ph]].

\bibitem{Bauer:2001ct}
C.~W.~Bauer and I.~W.~Stewart,
Phys. Lett. B \textbf{516}, 134-142 (2001)
[arXiv:hep-ph/0107001 [hep-ph]].

\bibitem{Bauer:2001yt}
C.~W.~Bauer, D.~Pirjol and I.~W.~Stewart,
Phys. Rev. D \textbf{65}, 054022 (2002)
[arXiv:hep-ph/0109045 [hep-ph]].

\bibitem{Bauer:2002nz}
C.~W.~Bauer, S.~Fleming, D.~Pirjol, I.~Z.~Rothstein and I.~W.~Stewart,
Phys. Rev. D \textbf{66}, 014017 (2002)
[arXiv:hep-ph/0202088 [hep-ph]].

\bibitem{Bauer:2002ie}
C.~W.~Bauer, A.~V.~Manohar and M.~B.~Wise,
Phys. Rev. Lett. \textbf{91}, 122001 (2003)
[arXiv:hep-ph/0212255 [hep-ph]].

\bibitem{Bauer:2003di}
C.~W.~Bauer, C.~Lee, A.~V.~Manohar and M.~B.~Wise,
Phys. Rev. D \textbf{70}, 034014 (2004)
[arXiv:hep-ph/0309278 [hep-ph]].

\bibitem{Manohar:2003vb}
A.~V.~Manohar,
Phys. Rev. D \textbf{68}, 114019 (2003)
[arXiv:hep-ph/0309176 [hep-ph]].

\bibitem{Becher:2006qw}
T.~Becher and M.~Neubert,
Phys. Lett. B \textbf{637}, 251-259 (2006)
[arXiv:hep-ph/0603140 [hep-ph]].

\bibitem{Goerke:2017ioi}
R.~Goerke and M.~Luke,
JHEP \textbf{02}, 147 (2018)
[arXiv:1711.09136 [hep-ph]].

\bibitem{Becher:2005pd}
T.~Becher and M.~Neubert,
Phys. Lett. B \textbf{633}, 739-747 (2006)
[arXiv:hep-ph/0512208 [hep-ph]].

\bibitem{Hoang:2008fs}
A.~H.~Hoang and S.~Kluth,
[arXiv:0806.3852 [hep-ph]].

\bibitem{Jouttenus:2011wh}
T.~T.~Jouttenus, I.~W.~Stewart, F.~J.~Tackmann and W.~J.~Waalewijn,
Phys. Rev. D \textbf{83}, 114030 (2011)
[arXiv:1102.4344 [hep-ph]].

\bibitem{Kelley:2011ng}
R.~Kelley, M.~D.~Schwartz, R.~M.~Schabinger and H.~X.~Zhu,
Phys. Rev. D \textbf{84}, 045022 (2011)
[arXiv:1105.3676 [hep-ph]].

\bibitem{Monni:2011gb}
P.~F.~Monni, T.~Gehrmann and G.~Luisoni,
JHEP \textbf{08}, 010 (2011)
[arXiv:1105.4560 [hep-ph]].

\bibitem{Boughezal:2015eha}
R.~Boughezal, X.~Liu and F.~Petriello,
Phys. Rev. D \textbf{91}, no.9, 094035 (2015)
[arXiv:1504.02540 [hep-ph]].

\bibitem{Moult:2018jzp}
I.~Moult and H.~X.~Zhu,
JHEP \textbf{08}, 160 (2018)
[arXiv:1801.02627 [hep-ph]].

\bibitem{Lin:2001zz}
C.~Lin, F.~H.~Zong and D.~M.~Ceperley,
Phys. Rev. E \textbf{64}, 016702 (2001)
[arXiv:cond-mat/0101339 [cond-mat.stat-mech]].

\bibitem{Sachrajda:2004mi}
C.~T.~Sachrajda and G.~Villadoro,
Phys. Lett. B \textbf{609}, 73-85 (2005)
[arXiv:hep-lat/0411033 [hep-lat]].

\bibitem{Bedaque:2004kc}
P.~F.~Bedaque,
Phys. Lett. B \textbf{593}, 82-88 (2004)
[arXiv:nucl-th/0402051 [nucl-th]].

\bibitem{Briceno:2013hya}
R.~A.~Briceno, Z.~Davoudi, T.~C.~Luu and M.~J.~Savage,
Phys. Rev. D \textbf{89}, no.7, 074509 (2014)
[arXiv:1311.7686 [hep-lat]].

\bibitem{10.2307/2033649}
H. F. Trotter,
Proceedings of the American Mathematical Society 10 (1959) 545.

\bibitem{Suzuki1976}
Masuo Suzuki,
Communications in Mathematical Physics 51 (1976) 183.

\bibitem{1976PThPh..56.1454S}
Masuo Suzuki,
Progress of Theoretical Physics 56 (1976) 1454.

\bibitem{kitaev2009wavefunction}
A.~Kitaev and W.~A.~Webb
[arXiv:0801.0342 [quant-ph]].

\bibitem{Qiskit}
H.~Abraham et.~al.~
10.5281/zenodo.2562110

\bibitem{ReadoutCorrection}
B.~Nachman, M.~Urbanek, W.~A.~de Jong, C.~W.~Bauer, 
npj Quantum Inf 6, 84 (2020)
[arXiv:1910.01969 [quant-ph]].

\bibitem{He:2020udd}
A.~He, B.~Nachman, W.~A.~de Jong and C.~W.~Bauer,
Phys. Rev. A \textbf{102}, no.1, 012426 (2020)
[arXiv:2003.04941 [quant-ph]].

\bibitem{futureKW}
C.~W.~Bauer, P.~Deliyannis, M.~Freytsis, B.~Nachman,
\emph{forthcoming}.

\bibitem{DAgostini:1994fjx}
G.~D'Agostini,
Nucl. Instrum. Methods Phys. Res. A \textbf{362} (1995) 487.

\bibitem{1974AJ.....79..745L}
L.~B.~Lucy,
Astron. J. \textbf{79} (1974) 745.

\bibitem{Richardson:72}
W.~H.~Richardson,
J. Opt. Soc. Am. \textbf{62} (1972) 55.

\bibitem{2010.07496}
R.~Hicks, C.~Bauer, and~B. Nachman,
Phys. Rev. A \textbf{103}, 022407 (2021)
[arXiv:2010.07496 [quant-ph]].

\bibitem{geller_efficient_2020}
M.~Geller and M.~Sun,
[arXiv:2001.09980 [quant-ph]].

\bibitem{song_10-qubit_2017}
C.~Song et al.,
Phys. Rev. Lett. \textbf{119} (2017) 180511.

\bibitem{gong_genuine_2019}
M.~Gong et al.,
Phys. Rev. Lett \textbf{122} (2019) 110501.

\bibitem{hamilton2020scalable}
K.~E.~Hamilton et al.,
[arXiv:2006.01805 [quant-ph]].

\bibitem{wei_verifying_2020}
K.~X.~Wei et al.,
Phys. Rev. A \textbf{101} (2020) 032343,
[arXiv:1905.05720 [quant-ph]].

\bibitem{pyquil}
Rigetti Forest,
http://docs.rigetti.com, 2020.

\bibitem{cirq}
The Cirq Contributors,\\
https://github.com/quantumlib/Cirq, 2020.

\bibitem{arute2020quantum}
F. Arute et al.,
[arXiv:2004.04197 [quant-ph]].

\bibitem{xacc}
The XACC Contributors,
https://xacc.readthedocs.io, 2020.

\bibitem{alex2019xacc}
A.~J.~McCaskey et al.,
[arXiv:1911.02452 [quant-ph]].

\bibitem{mccaskey_quantum_2019}
A.~J.~McCaskey et al.,
npj Quantum Information \textbf{5} (2019) 99.

\bibitem{ignis}
IBM Research,
https://qiskit.org/ignis, 2019.

\bibitem{1904.11935}
Y.~Chen, M.~Farahzad, S.~Yoo, and T.~Wei,
[arXiv:1904.11935 [quant-ph]].

\bibitem{1907.08518}
F.~B.~Maciejewski, Z.~Zimboras, and M.~Oszmaniec,
[arXiv:1907.08518 [quant-ph]].


\bibitem{Dumitrescu:2018}
E.~F.~Dumitrescu et al.,
Phys. Rev. Lett. \textbf{120} (2018) 210501.

\bibitem{PhysRevX.8.031027}
S.~Endo, S.~C.~Benjamin, and Y.~Li,
Phys. Rev. X \textbf{8} (2018) 031027.

\bibitem{PhysRevLett.119.180509}
K.~Temme,  S.~Bravyi, and J.~M.~Gambetta,
Phys. Rev. Lett. \textbf{119} (2017) 180509.



\end{thebibliography}
\end{document}